\documentclass[sigconf]{acmart}
\usepackage[utf8]{inputenc}

\usepackage[a-2b]{pdfx}
\usepackage{balance}
\usepackage[flushleft]{threeparttable}
\usepackage{multirow}
\usepackage{svg}

\newcommand{\circled}[1]{\raisebox{.5pt}{\textcircled{\raisebox{-.9pt}{#1}}}}

\AtBeginDocument{%
  }

\acmDOI{XXXXXXX.XXXXXXX}
\acmConference[SIGMOD '26]{International Conference on Management of Data}{May 31--June 05, 2026}{Bengaluru, India}
\title{GenJoin: Conditional Generative Plan-to-Plan Query Optimizer that Learns from Subplan Hints}

\author{Pavel Sulimov}
\email{pavel.sulimov@zhaw.ch}
\affiliation{
    \institution{Zurich University of Applied Sciences}
    \city{Winterthur}
    \country{Switzerland}
}
\authornote{Both authors contributed equally to this work.}

\author{Claude Lehmann}
\email{claude.lehmann@zhaw.ch}
\affiliation{
    \institution{Zurich University of Applied Sciences}
    \city{Winterthur}
    \country{Switzerland}
}
\authornotemark[1]

\author{Kurt Stockinger}
\email{kurt.stockinger@zhaw.ch}
\affiliation{
    \institution{Zurich University of Applied Sciences}
    \city{Winterthur}
    \country{Switzerland}
}

\renewcommand{\shortauthors}{Sulimov et al.}

\begin{document}

\begin{abstract}
    Query optimization has become a research area where classical algorithms are being challenged by machine learning algorithms. At the same time, recent trends in learned query optimizers have shown that it is prudent to take advantage of decades of database research and augment classical query optimizers by shrinking the plan search space through different types of hints (e.g. by specifying the join type, scan type or the order of joins) rather than completely replacing the classical query optimizer with machine learning models. It is especially relevant for cases when classical optimizers cannot fully enumerate all logical and physical plans and, as an alternative, need to rely on less robust approaches like genetic algorithms.
    However, even symbiotically learned query optimizers are hampered by the need for vast amounts of training data, slow plan generation during inference and unstable results across various workload conditions.
    In this paper, we present GenJoin - a novel learned query optimizer that considers the query optimization problem as a generative task and is capable of learning from a random set of subplan hints to produce query plans that outperform classical optimizers. GenJoin is the first learned query optimizer that significantly and consistently outperforms PostgreSQL as well as state-of-the-art methods on two well-known real-world benchmarks across a variety of workloads using rigorous machine learning evaluations.
\end{abstract}

\begin{comment}
\begin{CCSXML}
<ccs2012>
   <concept>
       <concept_id>10002951.10002952.10003190.10003192.10003210</concept_id>
       <concept_desc>Information systems~Query optimization</concept_desc>
       <concept_significance>500</concept_significance>
       </concept>
 </ccs2012>
\end{CCSXML}

\ccsdesc[500]{Information systems~Query optimization}

\keywords{Query optimization, Generative AI}
\end{comment}

\maketitle
%========================================================================================================
\section{Introduction}
\label{sec:intro}

% Today's built-in query optimizers of relational database management systems (RDBMS), like e.g. Postgres, combine the result of cost model estimation with 

Query optimization remains an active area of research for learned query optimizers (LQOs). In recent years, increasingly sophisticated methods have been developed for both cardinality estimation (CE) \cite{reiner2023sample, MSCN_kipf2018learned, NeuroCard_yang2020neurocard, wu2020bayescard, hilprecht2019deepdb, zhu2020flat, wu2023factorjoin, liu2021fauce, hilprecht2022zero, zhao2022queryformer} and join order selection (JOS) \cite{DQ_krishnan2018learning, ReJOIN_marcus2018deep, Balsa_Yang2022, Neo_Marcus2019, heitz2019join, yu2022hybridqo, RTOS_yu2020, chen2023loger, zhu2023lero, chen2023leon, Bao_Marcus2022, xu2023coool, woltmann2023fastgres, anneser2023autosteer}. CE approaches typically use statistical and machine learning (ML) models to approximate multivariate distributions over database table attributes~\cite{li2021cardinality}. The resulting cardinality estimates serve as an input to the cost models of query optimizers. At the same time, JOS models are considered to be the "brain" of query optimizers, whose outputs are logical and physical query plans ~\cite{ding2024lqos}.

\begin{figure}[h!]
    \centering
    \includesvg[width=\linewidth, keepaspectratio]{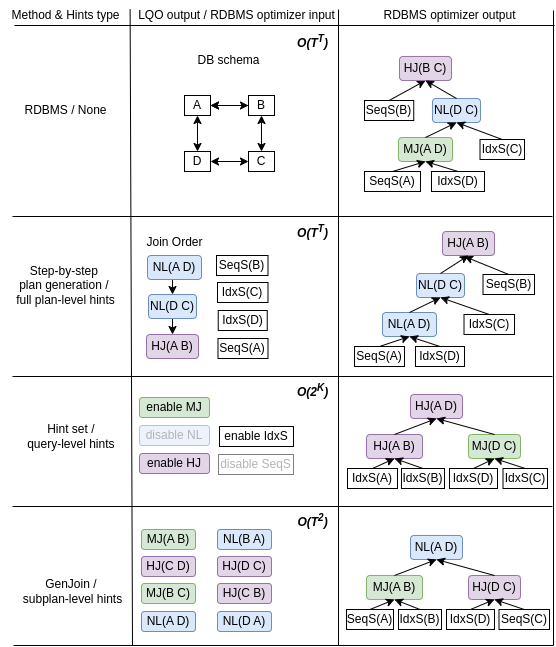}
    \caption{Illustration of how GenJoin is different from two main streams of learned query optimizers for evaluating the query $A \Join B \Join C \Join D$: step-by-step plan generation and hint set methods. NL: nested loop join, HJ: hash join, MJ: merge join, SeqS: sequential scan, IdxS: index scan, K: number of hints, T: number of tables.}
    \label{fig:genjoin_vs_others}
\end{figure}

Starting from Cascades~\cite{graefe1995cascades}, JOS was mainly considered as dynamic programming (DP)~\cite{cormen2009algorithms} task, naturally assuming a join choice as a step in top-down or bottom-up plan construction. With the rise of deep learning and reinforcement learning (RL)~\cite{sutton2018rl} as a logical continuation of the DP ideas~\cite{shengbo2023dprl}, classical query optimizers started being challenged by LQOs. Such step-by-step plan-building models control the full target plan specifications like the types of joins used, which scans to apply and in what order to join the tables, via a set of explicit hints for the RDBMS\footnote{Using extensions such as pg\_hint\_plan: \url{https://pg-hint-plan.readthedocs.io/en/latest/}.}. We call these approaches \textit{full plan-level hint}\footnote{The complexity of the potential prediction space is $O(T^T)$~\cite{wang1996complexity} scaling with the number of tables $T$.} methods (see second row of Figure~\ref{fig:genjoin_vs_others}). 

The trend of producing complete query plans as output was first questioned by Bao~\cite{Bao_Marcus2022}. The idea is to use classical optimizers and empower them by just giving high-level hints like 'enable merge\_join' rather than building complete query plans. In such a way, the exact join order choice is made by the built-in classical optimizer, though its search space is constrained through the provided \textit{query-level hints}\footnote{Each hint can be turned on or off independently, resulting in a potential prediction space complexity of $O(2^K)$ scaling with the number of hints $K$.} (see third row of Figure~\ref{fig:genjoin_vs_others}).

Despite the progress made, LQOs have yet to yield significant improvements upon traditional approaches and consistently outperforming them continues to be an elusive goal \cite{lehmann2024your}. We argue that the following weaknesses persist in today's LQOs:

\begin{enumerate}
    \item \textbf{Inefficient and limited RL.} Typical RL agents can take suboptimal steps and still end up with strong solutions due to the fact that either the game takes many steps to finish, and/or there is a possibility of stepping back. However, for query optimization and, in particular, bottom-up generation of join orders, every step is irreversible and has a significant impact on further steps with potentially fatal consequences in case of a bad choice. %Moreover, the described pipeline shrinks the plan search space, leaving an opportunity for only left-deep plans, even further dropping the chances of finding near-optimal solutions. %An example of such failure is the DQ method: it uses intermediate rewards of subplans to optimize the final execution time - though early isolated fast joins may lead to intermediate outputs that will result in slower further joins. %as the information gained by executing fatally flawed plans is limited at best. This effect is partially mitigated by methods like HybridQO, which applies Monte-Carlo Tree Search over join type hints.

    \item \textbf{Hint-based methods are double-edged swords.} Methods like Bao that give general hints (such as `disable\_hashjoin`, which impact every single pairwise join in a query) limit the query plan subspace like in k-d tree search~\cite{[panigraphy2008kdtree]}. This reduces the chance of finding the optimal solution due to the coarse level of granularity.
    %When recommending only general hint sets, methods like Bao leave a high degree of freedom to the classical optimizers to choose the order of joins. However, any hint, such as `disable\_hashjoin`, impacts every single join in a query, limiting the query plan subspace like in k-d tree search~\cite{[panigraphy2008kdtree]} at a coarse level of granularity and thus reducing the chance to find the optimal solution. % i.e. making the assumption that plan optimality is linear in hints space. 

    %\item \textbf{RL methods are not RL.} Many modern LQOs claim to be RL methods, and e.g. having replay buffer, value network (Neo), on-policy training strategy (Balsa). The thing is that those elements described are not obligatory RL, but might be simply "RL-inspired": value network is any regression neural network, replay buffer could be in GANs, on-policy is just using the information generated by the current network instead of sampling from the replay buffer, training with feedback loop can also be understood as active learning. Moreover, e.g. in Bao, it is claimed to use Thompson sampling~\cite{russo2018thompson} for solving the problem of hints set choice via contextual bandits~\cite{chapelle2011contextualthompson}. Though, as we can see from the official source code\footnote{https://github.com/learnedsystems/BaoForPostgreSQL}, vanilla deep contextual multi-armed bandits~\cite{collier2018contextualdeep} are used, having no update of the model regarding the regret and no stochasticity when choosing the arm.

    \item \textbf{Requirement for vast amounts of (training) data.} An important goal of query optimization is to learn the distributions and correlations of various attributes within and across tables to estimate the join cardinality. However, due to minute differences in query predicates, LQOs need to sample large amounts of query workloads to reach good generalization capabilities.
    
    \item \textbf{Finding optimal solutions with ML is time-intensive.} Certain ML algorithms do indeed find better query plans than traditional classical approaches. However, the time to find the solutions is often prohibitive due to the computational overhead for encoding queries or when using a model for inference which often greedily explores the plan space. Previous approaches often ignored these performance aspects and only focused on execution time as their only metric. 
    %While the execution time remains the primary objective for optimization, with increasingly growing machine learning models the time spent optimizing should be included as a performance metric.

\end{enumerate}

%Bao has too simple encoding that is not catching the structural specificities.

%For methods like Bao, we can suggest the use of Exp4 algorithm~\cite{bubeck2012exp4} instead of Thompson sampling, where the hint set is to be treated as an expert.

%Balsa is on-policy, but then how it solves the difference with update from Neo? Answer: Neo re-trains every time with all collected data (???)

As a possible way to mitigate these LQOs' hurdles, we developed GenJoin - a novel, \textit{generative plan-to-plan query optimizer that learns from subplan hints}\footnote{The complexity of the potential prediction space is $O(T^2)$ since all pairs of tables $T$ can participate in a subplan hint.} limited to join types such as nested loop, merge, and hash join. These three subplan hints are supported by the major database systems such as PostgreSQL, Oracle or SQL Server. 

The output of the GenJoin model is a \textit{set of two-way-join hints} (or subplan hints), e.g., use merge join on tables $A$ and $B$, i.e. $MJ(A,B)$, hash join on tables $C$ and $B$, i.e. $HJ(C,B)$, or nested loop join on tables $D$ and $A$, i.e. $NL(D,A)$. \textbf{GenJoin does not specify exactly where in the join tree this particular join should be performed (if performed at all)} but leaves it up to the classical optimizer. Also, this way, GenJoin gives the classical optimizer the freedom to choose all kinds of plans, including bushy ones. Moreover, GenJoin enables the classical optimizer to discover parts of the query plan space that it would not explore itself. For instance, the classical optimizer wanted to do $HJ(D,A)$ initially, but GenJoin recommends doing $NL(D,A)$ instead. This way, the classical optimizer might decide not to join $(D,A)$, which pushes it towards initially discarded alternatives.% to search for alternatives that were initially discarded.

\subsubsection*{Why does GenJoin recommend only the join type and neither the type of scans nor the join order?}

All the LQOs, including GenJoin, rely on internal RDBMS subquery cardinality estimations based on pre-calculated statistics. We never know if the RDBMS considers a sequential scan over an index scan due to high predicate selectivity. Moreover, we also do not know what the RDBMS has already cached, and LQOs typically ignore what is currently indexed and how. This applies that recommending just the join type gives the RDBMS the ability to solve those parts where it is more knowledgeable than we are: when performing a merge join, one table would require sorting, so it might be beneficial to perform an index scan.

We choose not to recommend the full join order but only the join type to avoid overfitting and to enable the LQOs to generalize better in a smaller search space.

\subsubsection*{Why is GenJoin a representative of the next generation of LQOs?}

In Table~\ref{tab:lqos_comparison} we present a comparison of GenJoin's architectural components against the state-of-the-art LQOs, as well as their predecessor LQOs. We emphasize the novelty of our method in terms of the following three categories:
\begin{itemize}
    \item \textbf{Plan Hints}: GenJoin suggests the "golden middle" between forcing exact query plans and giving a set of binary query-level hints (see bottom row of Figure~\ref{fig:genjoin_vs_others}).
    \item \textbf{ML Model}: GenJoin uses a conditional generative model with variational inference to create a region in the latent space from where an improved plan can be sampled.
    \item \textbf{Output Plan}: Sampling from a region implies the need to generate multiple solutions and to apply a filtering step. For GenJoin, it suffices to take a single sample, since the formation of the region already performs an \textit{implicit pruning}.
\end{itemize}

\begin{table}[h!t]
    \caption{Main architectural components of LQOs. Methods follow the chronological top-down order from oldest to newest.}
    \label{tab:lqos_comparison}
    \resizebox{\columnwidth}{!}{%
        \begin{threeparttable}
            
                \centering
                \begin{tabular}{l|ccc|cccc|ccc}
                    \hline
                    \multirow{2}{*}{LQO}                  & \multicolumn{3}{c|}{Plan Hints}      & \multicolumn{4}{c|}{ML Model}                            & \multicolumn{3}{c}{Output Plan}                                                       \\ \cline{2-11} 
                                                          & \rotatebox[origin=c]{80}{Full-plan}  & \rotatebox[origin=c]{80}{Subplan}    & \rotatebox[origin=c]{80}{Query}      & \rotatebox[origin=c]{80}{Regression} & \rotatebox[origin=c]{80}{\begin{tabular}[c]{@{}c@{}}Reinforcement\\Learning\end{tabular}} & \rotatebox[origin=c]{80}{\begin{tabular}[c]{@{}c@{}}Learning-\\to-Rank\end{tabular}} & \rotatebox[origin=c]{80}{Generative} & \rotatebox[origin=c]{80}{Single}     & \rotatebox[origin=c]{80}{\begin{tabular}[c]{@{}c@{}}Multiple \\ w/ Sampling\tnote{1}\end{tabular}} & \rotatebox[origin=c]{80}{\begin{tabular}[c]{@{}c@{}}Multiple \\ w/ Pruning\tnote{2}\end{tabular}} \\ \hline
                    Bao~\cite{Bao_Marcus2022}             & ~          & ~          & \checkmark & \checkmark   & \checkmark  & ~            & ~            & \checkmark & ~                                                                        & ~                                                                       \\ \hline
                    Lero~\cite{zhu2023lero}               & \checkmark & ~          & ~          & ~            & ~           & \checkmark   & ~            & ~          & ~                                                                        & \checkmark                                                              \\ \hline
                    HybridQO~\cite{yu2022hybridqo}        & \checkmark & ~          & ~          & \checkmark   & \checkmark  & ~            & ~            & ~          & ~                                                                        & \checkmark                                                              \\ \hline
                    COOOL~\cite{xu2023coool}              & ~          & ~          & \checkmark & ~            & ~           & \checkmark   & ~            & ~          & ~                                                                        & \checkmark                                                              \\ \hline
                    FASTgres~\cite{woltmann2023fastgres}  & ~          & ~          & \checkmark & \checkmark   & ~           & ~            & ~            & \checkmark & ~                                                                        & ~                                                                       \\ \hline
                    AutoSteer~\cite{anneser2023autosteer} & ~          & ~          & \checkmark & \checkmark   & \checkmark  & ~            & ~            & ~          & ~                                                                        & \checkmark                                                              \\ \hline
                    \textbf{GenJoin (ours)}               & ~          & \checkmark & ~          & ~            & ~           & ~            & \checkmark   & ~          & \checkmark                                                               & ~                                                                       \\ \hline
                \end{tabular}
                \begin{tablenotes}
                    \item[1] The method defines a region in the latent space where multiple plans could be selected.
                    \item[2] The method produces a number of intermediate plans during inference and afterwards applies a pruning algorithm to choose a single output plan.
                \end{tablenotes}
            
        \end{threeparttable}
    }
\end{table}

The \textbf{major contributions} of our paper are as follows:

\begin{itemize}
    \item We introduce \textit{GenJoin, a generative conditional query optimizer} that learns from only three types of subplan hints (i.e. nested loop join, merge join and hash join) using a generative model with variational inference-inspired architecture. GenJoin enhances query plans by \textit{pruning the search space of join types} of the built-in optimizer to greatly boost their query execution performance.
    \item We show, for the first time, that a \textit{learned query optimizer consistently outperforms} PostgreSQL as well as state-of-the-art methods on the two well-known real-world benchmarks JOB and STACK across a variety of workloads using rigorous ML evaluations.
    \item GenJoin is not only optimized for producing plans with low execution times but also \textit{minimizes the required inference time of the ML models} during query execution.
    \item Inside GenJoin, we introduce a \textit{new way of measuring the distance between query plans} via unnormalized difference of execution times and the corresponding p-value of a T-test for the means of plan execution time samples, which can be both used for training purposes and for calculating the confidence intervals of the query optimizers' differences.
\end{itemize}

\section{GenJoin Overview}

\subsection{General Setup}
\label{sub:genjoin_general}

%We aim to overcome the limitations outlined in the previous sections while taking advantage of current trends. Creating a plan from scratch (when only having information about the query and the database schema encoding) might be computationally expensive for training full plan-level hint generation models as they explore the whole plan space, i.e. require a huge amount of training data. The same holds true for the complexity of inferencing query-level hints when the number of possible hints is big enough. 

The basic idea of our approach is to start from some random query plan %\footnote{Naturally, it would be possible to feed in the plan generated by a built-in optimizer or another LQO. However, this would incur a substantial amount of overhead.} 
 (represented via a set of hints\footnote{Hereafter, we consider sets of subplan hints and query plans to be interchangeable, under the assumption that PostgreSQL turns a set of subplan hints deterministically into a query plan under static conditions.}) and train an ML model to improve the initial query plan using certain conditions (like the context of the query) - instead of a blind initial search. The research question we address is as follows: \textit{“How can we modify a set of hints such that the resulting query plan executes faster than the initial?”}

In order to be confident that the generated query plan results in an execution time that is not only faster than a random query plan, but also faster than the one produced by a built-in optimizer, we might need to apply our model multiple times on its own output. Such an architecture would be prone to overfitting (similar to other LQO methods that use model chaining for candidate plan selection and pruning). Instead, our model is designed to directly generate a query plan that outperforms PostgreSQL, which, as for now, is still state-of-the-art \cite{lehmann2024your}. For that, we need to incorporate additional auxiliary information, revealing the next question: \textit{“By how much is the target plan faster than the one produced by PostgreSQL?”}.

Combining these two questions, we find the answers with GenJoin - a \underline{gen}erative hint-to-hint method with the goal of providing a hint set (rather than fully specifying the join order), which leads to a faster query plan compared to a given plan and the plan produced by the baseline method, namely PostgreSQL.

\begin{figure}[h!t]
    \centering
    \includesvg[width=\linewidth,keepaspectratio]{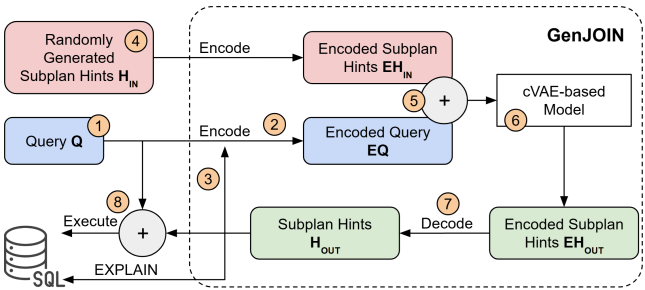}
    \caption{GenJoin query processing: Illustration of GenJoin and how it interacts with the RDBMS when a query is executed. The model is based on the conditional generative architecture with variational inference and produces an improved set of subplan hints $H_{OUT}$, given a query $Q$ and an initial set of subplan hints $H_{IN}$.}
    \label{fig:genjoin_overview}
\end{figure}

Figure \ref{fig:genjoin_overview} depicts the individual components of GenJoin for executing a previously unseen query Q \circled{1}. If running without GenJoin, PostgreSQL would execute the query using plan $P_{PG}$. The query $Q$ is encoded by querying the RDBMS for cardinality estimates of the filter predicates in $Q$ by individually querying each table using EXPLAIN \circled{2}. 
Next, we randomly generate an input plan $P_{IN}$ \circled{3} based on the query \circled{1} and transform it into a set of subplan hints \circled{4}. The set of subplan hints $H_{IN}$ is then encoded as $EH_{IN}$ for the generative model \circled{6}. 
The condition of our model \circled{5} is a combination of the encoded query $EQ$ and two p-values that represent the improvement of the output plan $P_{OUT}$ compared to the input plan $P_{IN}$ \circled{3} as $I_{IN}$ and compared to the PostgreSQL plan $P_{PG}$ \circled{1} as $I_{PG}$. During inference, the improvement values are set to zero. For additional information on the condition, see Section \ref{sub:condition}.

The encoded subplan hints $EH_{IN}$ (namely, \textit{nested loop join, merge join or hash join} of a given pair of tables) are concatenated with the condition \circled{5} as the input to our generative model \circled{6}. With the output of the encoder, we perform variational inference and pass the output and the condition to the decoder \circled{7}. GenJoin predicts the improved subplan hints $EH_{OUT}$ which are decoded to a set of subplan hints $H_{OUT}$ \circled{8}, so they can be added to the original query as query hints using \texttt{pg\_hint\_plan} and executed as plan $P_{OUT}$ \circled{9}.

%Figure \ref{fig:genjoin_overview} shows the components of GenJoin, which is inspired by the conditional variational auto-encoder (CVAE) \cite{sohn2015learning_CVAE} architecture. It uses hints to specify which type of join should be used on a per-join basis. The input is made up of the query encoding and two improvement indicators as the condition, which is fed to both the encoder and decoder modules, and the input plan to the encoder.

During training, pairs of plans are used where we can measure the difference between the two plans and the improvement factors. This allows our model to learn the traits of a good plan and which operations are needed to significantly improve over an input plan. 

%The details about query and plan encodings are given in Section \ref{sub:genjoin_encoding}, while the machine learning architecture, training and prediction are discussed in Sections \ref{subsec:model_architecture} through \ref{subsec:prediction}.

\subsection{GenJoin Encoding Scheme}
\label{sub:genjoin_encoding}

% General query/plan encoding
A variety of query encodings has been used in LQOs \cite{DQ_krishnan2018learning, ReJOIN_marcus2018deep, RTOS_yu2020, Neo_Marcus2019, Bao_Marcus2022, Balsa_Yang2022}, with varying levels of complexity. A common practice is to split the encoding into two sections, the \textit{query} and the \textit{plan encoding}. The query encoding contains all information about which tables are involved in a query and what kind of filters are applied on which columns. The plan encoding specifies which tables are joined, what the order of joins is, and the types of join used. This type of encoding represents at each step of the join tree which tables have been joined so far and how, and what else is left to join.

\begin{figure*}[h!t]
    \centering
    \includesvg[width=.8\textwidth, keepaspectratio]{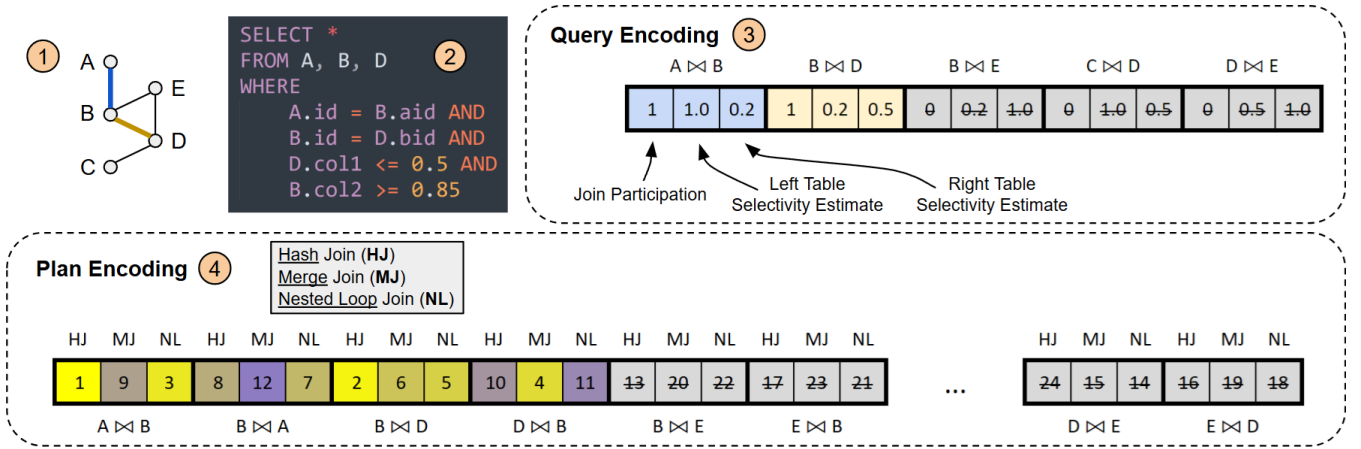}
    \caption{GenJoin encoding: \circled{1} Shows a simple data model depicted as a graph with tables A through E. Edges denote potential join paths. The join paths for the query A $\Join$ B $\Join$ D are shown in blue and yellow. \circled{2} Depicts the SQL query with join and filter predicates. \circled{3} Shows the \textit{query encoding}. For each potential 2-way join, the encoding marks the join participation and the estimated selectivity per table. The participating edges A-B and B-D are marked in \circled{1} and colored accordingly in \circled{3}. \circled{4} Shows the \textit{plan encoding} which indicates which join to perform. GenJoin ranks the bi-directional join options. All greyed-out cells with strikethrough text belong to non-participating joins for the example query.}
    \label{fig:genjoin_encoding}
\end{figure*}

% Database as graph
In case of GenJoin, the \emph{query encoding} is similar to the one used in Neo~\cite{Neo_Marcus2019} (and the majority of LQOs released after it). The \emph{plan encoding} is new and specific for the given subquery hints, i.e., this is the novelty coming along with the method itself. To illustrate our encoding scheme, we have prepared an example in Figure \ref{fig:genjoin_encoding}. Our proposed query encoding views a database as a graph (see \circled{1}), where each table is a node and every edge indicates a potential join path between two nodes. In the example, we can see that the five tables A through E have different ways to be joined, e.g., one can join tables A and B directly (indicated by the blue edge), but not B and C. As an example, we have prepared a query (see \circled{2} in Figure \ref{fig:genjoin_encoding}) joining tables A, B and D with predicates on tables B and D. Assuming the columns B.col2 and D.col1 are uniformly distributed between $0$ and $1$, they result in a selectivity of $50\%$ and $15\%$ for those two tables, respectively. 

% Query Encoding
The \textit{query encoding} (see \circled{3} in Figure \ref{fig:genjoin_encoding}) contains one 3-sized cell for every edge in the graph of potential 2-way joins. In our example, that means there are five cells for the 2-way joins $A \Join B$, $B \Join D$, $B \Join E$, $C \Join D$ and $D \Join E$. Each cell contains one number to indicate whether this join participates in the query and two numbers corresponding to the estimated selectivity after all filter predicates are applied to this table. For example, the blue cell for $A \Join B$ contains $(1, 1.0, 0.2)$, since the join is part of the example query (see \circled{2} in Figure \ref{fig:genjoin_encoding}) resulting in a $1$, table A has no filters applied resulting in $1.0$ and table B with the filter $B.col2 >= 0.85$ resulting in an estimated selectivity of $0.2$. Please note, that the selectivities of the encoding are extracted estimates (for example, from EXPLAIN calls to PostgreSQL) and not the true selectivity (hence $0.2$ rather than the true $0.15$ in the example).

% Plan Encoding
The \textit{plan encoding} (see \circled{4} in Figure \ref{fig:genjoin_encoding}) uses the same idea of 3-sized cells like the query encoding, but encodes each edge in both directions (as there are significant differences in hinting $A \Join B$ versus $B \Join A$, e.g., depending on whether the tables are sorted or not). Unlike the query encoding, the plan encoding solely contains the information about \textit{join types and rankings} thereof. Each cell contains three entries for ranking the hash, merge, and nested loop joins on this particular join. In our example, lower values mean a higher rank e.g., 1 being first rank. The cell for $A \Join B$ contains the ranks $1$, $9$ and $3$ (ranked across all bi-directional join options), implying that GenJoin would force PostgreSQL to use a hash join if it were to join tables A and B in the order $A \Join B$.

One would assume from the example that PostgreSQL performed a hash join on $A \Join B$ (rank 1) and a hash join on $B \Join D$ (rank 2), since these are the highest ranking join types across all participating joins. However, GenJoin does neither fully specify the order in which the joins are executed nor in which order each pair of tables is to be joined. In our example, PostgreSQL could also decide to instead reverse either join as a nested loop join of $B \Join A$ (rank 7) and a merge join of $D \Join B$ (rank 4).

The idea behind the plan encoding is that GenJoin fully specifies all subplan hints provided to PostgreSQL by the highest ranking value in every cell. Moreover, individual rankings can also be understood as a confidence of the model in a particular subplan hint.

\subsection{Model Architecture: Conditional Generative Model with Variational Inference}
\label{subsec:model_architecture}
In this section, we describe the ML model architecture used in GenJoin. Our approach is inspired by conditional variational autoencoders (cVAE) \cite{sohn2015learning_CVAE}. We suggest an encoder-decoder architecture, include conditional information and apply variational inference as shown in Figure \ref{fig:genjoin_overview}. We use a set of subplan hints as our input (the plan encoding) and use the query encoding as our condition. Likewise, the condition contains auxiliary information on the expected performance of the generated output (see Section \ref{sub:condition} for more details). The output of our model is another set of subplan hints.

% \begin{figure}[h!t]
%     \centering
%     \includegraphics[width=\linewidth, keepaspectratio]{images/fig__genjoin_architecture.png}
%     \caption{\textcolor{blue}{Architecture of GenJoin using a conditional variational autoencoder.}}
%     \label{fig:gen_join_cvae}
% \end{figure}

\begin{figure}[h!t]
    \centering
    \includesvg[width=\linewidth, keepaspectratio]{images/fig__latent_sampling.svg}
    \caption{GenJoin inferencing: The encoder makes a projection to the region in the latent space where hint sets leading to faster query plans are presumably located. A hint set is generated by sampling from the region using variational inference, followed by the decoder, which turns the latent representation into a hint set.}
    \label{fig:latent_sampling}
\end{figure}

%The encoder-decoder architecture is not entirely new in the space of LQOs. For example, \cite{marcus2023learned} presented the idea of optimizing queries in the latent space.
Until now, the potential of using autoencoders for query representation was presented as an idea~\cite{marcus2023learned}, suggesting to optimize queries in the latent space. This way, the encoder is limited to only performing a dimensionality reduction by learning a \textit{bijective} function that maps query plans into a latent space where semantic properties emerge. We make the encoder in GenJoin perform a different task instead: Our novelty is that we learn an \textit{injective} projection of a \textit{set of subplan hints} into a region of the latent space from which we can directly sample a superior set of subplan hints (see Figure \ref{fig:latent_sampling}) in a single shot without the need to apply additional latent space optimization algorithms, which could be both expensive and prone to error propagation. The advantage of our approach is that during inference, GenJoin guides the RDBMS to turn the input set of subplan hints into a faster-running physical query plan through latent space injection (which performs implicit pruning of the latent space to find the optimal region).

\subsubsection*{Why can we expect to find a faster set of subplan hints given a random plan?} For any given input set of subplan hints there exist many sets of subplan hints (or consequently resulting query plans) that outperform the initial one, because the probability of randomly providing the fastest plan as input is low. These resulting sets of subplan hints form a clustered region in the latent space containing potentially faster query plans. Even if the optimal plan is provided as an input, GenJoin can provide sets of subplan hints that result in the same executed physical plan through dynamic optimization~\cite{hellerstein2000dynamic}.

%\subsubsection*{Why does this region always exist?}, and (2) different sets of subplan hints may result in the same generated plan through dynamic optimization~\cite{hellerstein2000dynamic}.} 

%The difference in GenJoin is that the encoder performs a different task, where the \textit{injective} projection of the input is positioned in the latent space of hint sets such that sampling from the region of hint sets results in a beneficial set of subplan hints (see Figure ~\ref{fig:latent_sampling}), leading to \textit{one of several faster-running plans}\footnote{We can reasonably assume that for any given input plan there will be many other plans that outperform it, forming regions in the latent space where such plans are located. This is because the probability of providing the optimal plan as input is low, and different subplan hints may result in the same generated plan through dynamic optimization~\cite{hellerstein2000dynamic}.}.
 
% cVAEs are an extension of the variational autoencoder (VAE) \cite{kingma2013auto_VAE}, which itself is a continuation of the autoencoder (AE) \cite{hinton2006reducing_AE} idea. AEs are an encoder-decoder architecture where the input is compressed into a vector in a lower-dimensional space using the encoder, which aims to reconstruct the input based on the latent representation using the decoder. 

\subsubsection{Model Design Details}

For GenJoin, the encoder is designed to map the input (the query and plan encodings, as well as the conditioning information about the desired output plan) to a distribution in the lower-dimensional latent space, typically assumed to be a Gaussian distribution (as the inputs are not randomly projected, but rather constrained to lie within a restricted subspace).
%which follows some pre-defined distribution (i.e., inputs are not randomly projected into a lower-dimensional space but are instead constrained to lie within a restricted subspace, typically a Gaussian distribution).
By sampling from this region, the decoder performs variational inference~\cite{blei2017vi}.

Training such a model is made possible by the "reparametrization trick"~\cite{fu2006reparametrization}. For example, if we decide to project our input into a standard normal latent space $N(0, 1)$, the encoder learns two parameter vectors $\mu$ and $\sigma$, aiming to keep the mean $\mu \approx 0$ and the standard deviation $\sigma \approx 1$. The decoder can randomly sample $\epsilon$ from $N(0, 1)$, add our $\mu$, and multiply by our standard deviation $\sigma$: 

\begin{equation} 
\label{eq:reparametrization_trick}
z = \mu + \sigma \bigodot \epsilon 
\end{equation} 

The result continues to follow the distribution $N(0, 1)$. With this trick, we have a differentiable Equation ~\ref{eq:reparametrization_trick}, allowing the gradient flow through the neural network since we delegate the random sampling to a noise vector, effectively decoupling it from the gradient flow.

\subsection{Training Data Generation}
\label{subsec:training_data_generation}

Since GenJoin uses subplan hints both for its input and output, we have to generate training data specifically for our model in the form of pairs of sets of subplan hints, which implies our approach is data-centric rather than model-centric. The challenge is that, learning to predict a "better" set of subplan hints, the pairs need to exhibit a significant difference in execution time and are constructed such that the output set of hints always results in a faster query execution compared to the input set of hints.

\begin{figure*}[h!t]
    \centering
    \includesvg[width=\textwidth, keepaspectratio]{images/fig__training_data_generation_new.svg}
    \caption{GenJoin training data generation: We show three different sets of subplan hints $H_1$, $H_2$ and $H_3$ obtained from a randomly generated plan for query $A \Join B \Join C \Join D$. The sets are constructed from the physical operators of the given plan (such as hash join for $D \Join C$) as a hint set template. The remaining combinations of pairwise join types that are not present in the plan are chosen arbitrarily. The PostgreSQL optimizer, restricted by $H_i$, produces a physical plan $P_i$, that is executed 3 times. From these execution time measurements, we calculate the pairwise p-values of queries executed with hint sets $H_i$ and $H_j$. Pairs with significant differences are selected for training. For example, we only keep the pair $<(H_1, condition), H_2>$ where $(H_1, condition)$ is the input and $H_2$ the output of our model.}
    \label{fig:training_data_example}
\end{figure*}

% \begin{figure}[h!]
%     \centering
%     \includegraphics[width=\linewidth, keepaspectratio]{images/fig__training_data.png}
%     \caption{Example of three different sets of subplan hints $H_1$, $H_2$ and $H_3$ for query $A \Join B \Join C \Join D$ with the training pairs based on their execution time $e_i$. Note that the subplan hints are bi-directional, e.g., for the sub query A $\Join$ B the subplan hints are MJ (A,B) and MJ (B,A) for $H_1$. For $H_2$ we see a different hint set of MJ (A,B) and NL (B,A).}
%     \label{fig:training_data_example}
% \end{figure}

For every query in the training set, we randomly generate 200 sets of subplan hints $H_i$ and pass them into PostgreSQL to measure the corresponding execution time $e_i$. It is worth noticing that generating training data in such a way, i.e. treating PostgreSQL as a black-box environment, sending a set of hints, and receiving feedback in the form of a query plan, aligns with the concept of direct RL, which was advised in \cite{lehmann2024your}. This approach exempts us from describing the environment, i.e. knowing the details of the internal PostgreSQL rules, but rather gives the possibility to learn how it reacts to a given set of subplan hints. See Figure \ref{fig:training_data_example} for an example with 3 sets of subplan hints for the query A $\Join$ B $\Join$ C $\Join$ D.

We generate pairs of sets of subplan hints, which are combined with the global context of the query encoding and the condition, yielding tuples in the form of $<(H_i, condition), H_j>$. $H_i$ represents the slower of the two sets of subplan hints and is fed to the model as input together with the condition (see Section~\ref{sub:condition} for more details), while $H_j$ is the faster set of subplan hints representing the desired output. Since we are interested in learning a function that improves a given set of subplan hints, we only keep pairs for which there is a statistically significant difference between the execution times $e_i > e_j$. For this, we require a metric that provides us with a level of confidence for the difference between execution times and measures the distance between any two queries using sets of subplan hints. We conclude that the best choice is a p-value.

\subsubsection*{Why choose a p-value as a proxy for the distance between arbitrary query plans?}
\label{sec:p_value}
% Absolute distance is bad, |e1-e2| different for fast and slow queries in terms of position for ML model learning effectiveness and instability

First of all, since we know from \cite{lehmann2024your} that the true execution time $e_i$ cannot be exactly measured due to cache states and similar database properties, we use the mean estimation $\overline{e_i}$ calculated over a number of executions. Thus, measuring the unnormalized difference $\overline{e_i} - \overline{e_j}$ gives a distance metric, though it can produce values in an unbounded range from query to query. This may lead to learning inefficiencies and training instability for ML models.
% relative distance is good, but unclear how to normalize (stddev?) to mitigate the effect of scale
% also relative follow the learing to rank trends
LQOs based on learning-to-rank models~\cite{chen2023leon, zhu2023lero, xu2023coool} utilize relative rankings instead, solving the ML issue mentioned above, though they lose the ability to measure how similar plans are.

One option to force the values of the unnormalized difference to be inter-query comparable is Studentization~\cite{kendall1973statistics} - a division of a first-degree statistic derived from a sample by a sample-based estimate of a standard deviation - creating a scale-free metric. For the difference of means, such a normalization factor could be a pooled standard deviation $s_p$, i.e. a compound of standard deviations $s_{e_i}$ and $s_{e_j}$ with corresponding sample sizes $n_i$ and $n_j$:

\begin{equation} 
\label{eq:standard_deviation_pooled}
s_p = \frac{1}{\sqrt{\frac{{s_{e_i}}^2}{n_i} + \frac{{s_{e_j}}^2}{n_j}}}
\end{equation}

The series of mathematical manipulations described above end up with nothing else but an empirical T-test~\cite{student1908ttest} statistic for the means of the two independent query plan execution time samples. To achieve the property of distance significance, we compute the p-value associated with the created empirical T-test statistic:

\begin{equation} 
\label{eq:p_value}
\text{p-value}(\frac{\overline{e_i} - \overline{e_j}}{s_p})
\end{equation}

%The ideal normalization factor is open for discussion but should be proportional to both execution times, as we cannot guarantee which of two arbitrary query plans results in a larger execution time. 
% Requirement for multiple measurements to get empirical distributions (refer to paper)
To ensure having only statistically significant differences, we keep only pairs with p-value $\leq$ 0.025, i.e. confidence of 5\% for the one-tailed hypothesis. The p-value serves as a bounded metric for comparing the distance between queries. Being outlier-robust, the p-value enables ML models to learn from timeout queries.
%However, this requires us to sample multiple measurements to generate an empirical distribution of execution times. 
%Because of this, we discard the 1st and 2nd execution time, but how many additional measurements to take beyond the 3rd is discussed in the next section.

%Choosing the number of samples $N$ to take is a tradeoff between speed and accuracy.  We arbitrarily choose $N=3$, meaning our empirical distribution of execution time is made up of the 3rd, 4th and 5th execution.
% voilà p-values of the t-test to the rescue, implicit standardization

%From our previous article~\cite{lehmann2024your}, it is clear that starting from the third consecutive execution, we can trust the runtimes as an unbiased estimation of the true execution time. However, just taking a stand-alone third execution time result is not sufficient because

\subsubsection*{How many samples should we choose for the p-value measurements?}
The p-value can only be measured between the distributions of query execution times, i.e. we need to collect $\ge 2$ query samples from the RDBMS. Since we do not have a predefined minimal detectable effect nor a desired statistical power, we choose the number of query samples arbitrarily, aiming to have a stable variance with the minimal executions-per-query required. Experimentally studying samples from standard normal distributions of different sizes and applying the elbow rule~\cite{thorndike1953elbow}, we conclude that starting from \textit{sample\_size=3}, the change in variance is no longer significant.

This way, for each hint set pair, we compute the p-value of the T-test between the input and output execution times. We do the same between the output execution time $e_{out}$ and the execution time of our baseline method (PostgreSQL) $e_{pg}$. Starting with the 3rd consecutive execution, we can trust the runtimes as an unbiased estimation of the true execution time~\cite{lehmann2024your}. Our empirical distribution is made up of the 3rd, 4th and 5th query execution.

\subsubsection*{How can the unnormalized difference be used for comparing LQOs?}

To make sure that one LQO has an advantage over the other in terms of execution time on a fixed workload of queries, we can utilize the unnormalized difference introduced above, using the pooled standard deviation $s_p$ from Equation~\ref{eq:standard_deviation_pooled} for computing the confidence intervals aka error bars. For relative differences, it would be impossible to provide these kinds of error bars. One may ask, why using the approach for a single query difference can be extrapolated from a set of queries in a workload. We argue that as long as for the two normal distributions~\cite{lemons2002sumofnormals} $X$ and $Y$, the following is true:

\begin{equation*}
    \begin{split}
    X\sim N(\mu _{X},\sigma _{X}^{2}), Y\sim N(\mu _{Y},\sigma _{Y}^{2}) \Rightarrow \\
    Z=X+Y \sim N(\mu _{X}+\mu _{Y},\sigma _{X}^{2}+\sigma _{Y}^{2})
    \end{split}
\end{equation*}

and our execution time samples per query are assumed to be normally distributed, the methodology described in Section~\ref{sec:p_value} above is generalizable for the sum of differences.

\subsubsection{Conditional Information in GenJOIN}
\label{sub:condition}

The conditional information allows us to specify what kind of output should be generated. To transform one set of subplan hints into another, we add supplementary contextual information through the query encoding (see Section \ref{sub:genjoin_encoding}), as well as information about the improvement in performance across the two hint sets. The level of improvement is captured with two distinct p-values: one approximates the difference between the performance of the query using the output set of subplan hints and the input set. The other p-value represents the difference in query performance between the output set of subplan hints and the set produced by PostgreSQL, answering the questions from Section \ref{sub:genjoin_general}, how much faster the target plan is, compared to both the input plan and the plan produced by PostgreSQL. During training, all plans and their execution times are available and thus all improvement p-values can be calculated. During inference, both p-values are set to the minimum value without the need to explicitly measure the performance of other plans.

\subsection{Prediction of Subplan Hints}
\label{subsec:prediction}

In the previous section, we outlined how the training data, and especially the pairs of input subplan hints $H_{IN}$ and output subplan hints $H_{OUT}$, are generated. For the prediction, we also choose a randomly generated initial set of subplan hints as $H_{IN}$ while our model predicts the subplan hints $H_{OUT}$ (as shown after step 7 in Figure \ref{fig:genjoin_overview}), which are then attached to the query and executed.

This leaves the question of what to assign to the p-values in the condition, namely for the confidence to surpass PostgreSQL and the confidence to surpass the input plan. Both of these values are set to zero, which corresponds to the theoretically best p-values under the hypothesis that the generated plan $H_{OUT}$ performs better than both PostgreSQL and the input plan $H_{IN}$.

However, these p-values only produce a plan that is potentially better than the one initially given, contrary to the general aim of query optimization to find the best possible plan. Since the suggested approach considers the output $H_{OUT}$ to be better than input $H_{IN}$, we can run the prediction in a loop - a so-called "chain-of-subplan-hints" - assuming that every next output is faster than the previous input (see Section \ref{sub:ablation_chain_of_plans} for the results of this ablation study). %This concept follows the chain-of-thoughts idea from prompting LLMs ~\cite{wei2022cof}, recently also applied to OpenAI's o1 model\footnote{\url{https://openai.com/index/learning-to-reason-with-llms/}}. 

%The main problem with this chaining approach is there is no infinite potential for improvement - we cannot know in advance how drastic the potential performance improvement is, and with that we take a risk that after $N$ iterations, the model starts giving some random hint sets. In order to mitigate this problem, we can introduce the protocol of early stopping. Our classification model, working after main output hint set generation, returns the evaluation of e.g. confidence of 

% ========================================================================================================
\section{Experiments and Results}

In this section, we provide our experimental evaluation of GenJoin on the Join Order Benchmark (JOB) and STACK. We start by describing the experimental setup, present our results of the experiments and finish with a number of ablation studies.

\subsection{General Setup}

\subsubsection{Software and Hardware}

Planning and execution times are measured using \texttt{EXPLAIN ANALYZE}. The LQOs' inference time is measured as the wall time for e.g., preprocessing, encoding or inferencing. Measurements are taken in a hot cache setting by executing the same query three times and taking the last query execution.

PostgreSQL is configured as in \cite{lehmann2024your}, with \texttt{AUTOVACUUM} disabled to keep the sampled statistics consistent across the experiment. An \texttt{ANALYZE} call was run after setting up each workload. The experiments were run on machines with 64 GB of RAM, 16 CPU cores and Tesla T4 GPUs using a Docker environment.

We emphasize that a large number of publications are based on PostgreSQL versions 12 \cite{Balsa_Yang2022, Bao_Marcus2022, chen2023leon, xu2023coool, lehmann2024your} and 13 \cite{anneser2023autosteer, zhu2023lero}. All our experiments were conducted using PostgreSQL version 16 to reflect the incremental changes added over time, further increasing the difficulty of outperforming the PostgreSQL baseline.

\subsubsection{Query Workload}

For our experiments, we evaluate all methods on two established query optimization workloads, namely the Join Order Benchmark (JOB) \cite{JOB_Leis2015} and the StackExchange-based STACK \cite{Bao_Marcus2022}. These two real-world benchmarks give good insights into the overall performance of various methods, including many queries that are hard to optimize \cite{JOB_Leis2015}. JOB has been the primary benchmark for query optimization in recent years, with STACK in the runner up position since its release in 2021. Both benchmarks show skewness in the data and violate the uniformity assumption.

\textbf{Join Order Benchmark (JOB).}
JOB is comprised of queries around a snapshot from May 2013 of the popular Internet Movie Database (IMDB). It contains 21 tables stored in a relational database. The queries have, on average, 8 joins per query with a maximum of 16 joins. In total there are 113 queries that come from 33 base queries (or templates) that each have between two and six variations, where either filter predicates are changed or entirely new attributes are being filtered on. These predicates have a significant impact on the selectivity of the tables involved. Hence, the optimal physical plan might differ from variant to variant. These differences in execution time can span multiple orders of magnitude. However, well-performing plans often share partial subplans with other variants, creating a potential for data leakage.
%Because the predicates have a significant impact on the selectivity of the tables involved, the optimal physical plan can be varied resulting in execution time differences of up to multiple orders of magnitudes. However, partial sub plans are often shared across variants for optimal sub plans and are informative in solving other query variants.

\textbf{STACK.}
Introduced alongside the Bao method \cite{Bao_Marcus2022}, the STACK workload and dataset is based on data from the StackExchange websites. STACK features 10 tables and over 6,000 queries from 16 base queries (or templates). The queries follow a similar variation style as in JOB, meaning there exist many different variations in the filter predicates for every base query available. As per \cite{lehmann2024your}, we have downsampled the numbers of queries again to keep the comparison between methods at a similar statistical power. For this reason, we use 112 queries with 7 variants for each of the 16 base queries.

\subsubsection{Training Data Generation}

For our workloads with 112 and 113 queries and 200 randomly generated hint sets per query, we can potentially generate up to 4.5 million pairs\footnote{$112 * 200 * 200 \approx 4.5$ million, though we are interested in an imposed order of their execution time, which leaves half that amount before checking significances.}. Due to the limitation of only selecting them in order of execution time and with significant differences, the resulting number of usable training pairs is on the order of about 1 million. We subsample the actual amount of pairs and evaluate the ratio of pairs for training using the Optuna hyperparameter optimizer\footnote{\url{https://optuna.org/}}. For JOB we use $30\%$ of all pairs during training, for STACK it is $45\%$. These ratios serve as a surrogate for classical early stopping mechanisms because our training data volume regulation is aimed to reduce the generalization error~\cite{bishop2006generalization}. In line with other LQOs, supporting a new database requires generating new training data and train a new model.

\subsubsection{Dataset Split}

\begin{figure}[h!]
    \centering
    \includesvg[width=\linewidth, keepaspectratio]{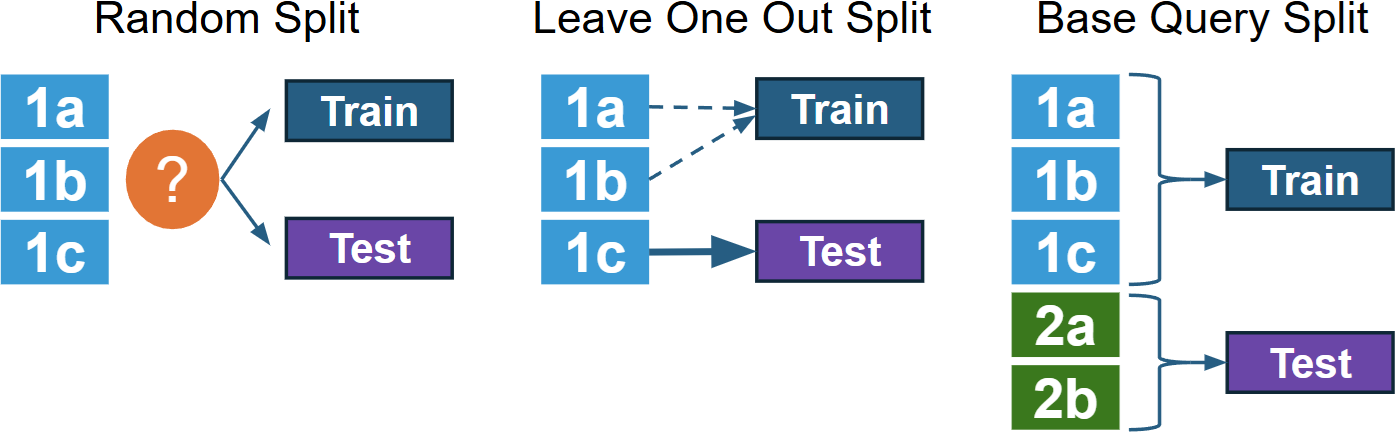}
    \caption{Summary of the dataset split types used in order: Random Split, Leave One Out Split and Base Query Split.}
    \label{fig:split_types}
\end{figure}

We continue using the different types of splits described in \cite{lehmann2024your}, namely the Random, Leave One Out and Base Query split types. These dataset splits take advantage of the structure of the query workloads, where queries can be clustered together as having different filter predicates, but an overall shared structure. These shared structures are called \textit{base queries}, with the different variations of the filter predicates being called base query variants. Figure \ref{fig:split_types} gives a summary for how the queries 1a, 1b and 1c belonging to the same base query 1 would be assigned to the train and test set. In the \emph{random split}, queries are randomly assigned to the train and test set. In the \emph{leave one out split}, all variants of a base query except one are put into the train set. Finally, the \emph{base query split} assigns all base query variants either to the train or test set, respectively. Hence, queries 1a, 1b and 1c end up in the train set together, while queries 2a and 2b are collectively put into the test set. While the \emph{base query split} aims to reduce data leakage across shared query structures, the \emph{leave one out split} forces a minimum level of shared information between the training and test sets.

The various splits emphasize different characteristics of the query workloads and give strong overall insights into the models' performances. We have made one adjustment compared to \cite{lehmann2024your}: For every type of split, we distribute the data according to the split function into three separate folds in the style of k-fold cross-validation. The three splits are then called A, B and C, e.g., "Base Query Split A", "Base Query Split B", and "Base Query Split C", where the \textit{test sets together form partitions of all queries}. Since the test sets of splits A, B and C form a congruent set of all queries, we list the performance of a model on that type of split as the combination of all three test sets. This allows us to make a more accurate, direct comparison across split types.

There are two special cases that need to be handled: First, if a group of the same base query has a number of variants that is not divisible by three, the remainder is randomly assigned between the three folds. Second, if there are fewer than three query variants, some variants will be assigned to multiple splits. In the latter case, the evaluation takes the worse of the two query's measurements. 

For a number of ablation studies, we also use the \emph{slow split}, first introduced by \cite{Balsa_Yang2022} for JOB as \emph{JOB-Slow}. This takes the slowest $N$ queries (for JOB-Slow in Balsa $N=19$) into the test set and maximizes the absolute amount of time that can be optimized in query executions. The extension of the slow split for STACK follows the same notion and includes the 19 slowest queries in the test set.

% TODO: Do we need to describe the ANALYZE state of different machines?

%What kind of metrics we used for evaluation and why we measure only relatively to Postgres with some relative difference.

%A new approach differs not only in terms of splits forming (A, B, C) but also because it looks at differences explicitly with confidence intervals (and potentially with aggregating on splits, and making relative and absolute differences comparison at same time).

%Exponentially weighted MAPE is required not only because absolute values are not comparable between machines and databases, and relative can give biased results because of speeding up already fast queries - but because also depending on initial sampling during ANALYZE, we can have very different distribution of execution times from Postgres (which also influences training). So that, we can introduce weighted percentage (with softmax) difference as a final measure.

%We can compare between each other methods trained on different machines and under different ANALYZE, because we only make comparison to PG on the same machine, and final metrics is to be relative.

%Now running with hot cache is even more important because it might be the problem that same set of hints is given to Postgres, but it executes it differently under different cache etc.

\subsection{Comparing GenJoin against Current State-of-the-Art Learned Query Optimizers}
\label{sub:comparison_state_of_art}
% This section includes the "normal" experiments, where we show that we outperform JOB/STACK without looking at inference time) and without looking at SLOW

In this section, we compare the performance of GenJoin with two state-of-the-art LQO methods - whose source code is available, run with PostgreSQL Version 16, and the results are reproducible - namely, HybridQO \cite{yu2022hybridqo} and AutoSteer \cite{anneser2023autosteer}. We have chosen HybridQO since it outperformed Neo \cite{Neo_Marcus2019}, Balsa \cite{Balsa_Yang2022}, Bao \cite{Bao_Marcus2022} and LEON \cite{chen2023leon} in previous evaluations \cite{lehmann2024your}.

Bao has been extensively used for comparison in recent LQO publications, but because it does not natively run under PostgreSQL version 16, we needed an alternative for the query-level hint methods. While both AutoSteer and FASTgres \cite{woltmann2023fastgres} improve upon Bao, AutoSteer adds a dynamic exploration of tuning knobs in addition to a much larger number thereof, while FASTgres primarily uses a different type of model with the same number of knobs as Bao. Therefore, we have chosen AutoSteer for our evaluation, which has a more expressive optimization space over FASTgres and Bao.

We have also considered including COOOL \cite{xu2023coool}, but the experiments could not be reproduced\footnote{We have contacted the authors of COOOL because their published code only includes model checkpoints and the code for inference, but lacks the required scripts for training a model. The authors did not provide the necessary training code, under the motivation that "posting the training code may confuse readers because the results in the paper cannot be obtained via re-training".} - similar to other non-reproducible methods discussed in \cite{lehmann2024your}. For additional details on other LQO methods, we refer to our Related Works in Section \ref{sec:related_work}.

% TODO Discussion about which times to use

\begin{figure}[h!]
    \centering
    \includesvg[width=\linewidth, keepaspectratio]{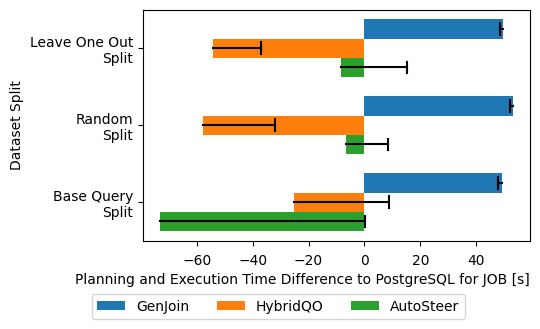}
    \caption{Results on the Join Order Benchmark (JOB) without inference times. Methods are compared against the PostgreSQL baseline. If the bars extend into the positives, the methods are faster than PostgreSQL and vice versa.}
    \label{fig:results_job__no_inference}
\end{figure}

\begin{figure}[h!]
    \centering
    \includesvg[width=\linewidth, keepaspectratio]{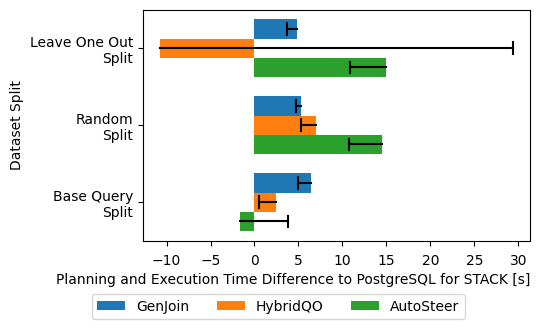}
    \caption{Results on STACK without inference times.}
    \label{fig:results_stack__no_inference}
\end{figure}

% Order of discussing splits
Figures \ref{fig:results_job__no_inference} and \ref{fig:results_stack__no_inference} show the performance of GenJoin compared against HybridQO and AutoSteer executed on JOB and STACK, respectively. The methods are evaluated along the three types of splits outlined in the previous sections, in order of difficulty, starting with the leave one out split, then the random split and finally discussing the base query split. Please note that \textbf{the level of HybridQO and AutoSteer performance is different compared to the one stated in original papers}~\cite{yu2022hybridqo, anneser2023autosteer} due to the usage of fair training and testing procedures justified in \cite{lehmann2024your}.

% Explain how the figure is read
% TODO Explain what the statistical significance is
In Figures \ref{fig:results_job__no_inference} and \ref{fig:results_stack__no_inference} we compare the methods against PostgreSQL since executing queries faster than the classical reference is the primary goal. The x-axis describes the difference in \emph{planning and execution time} summed over all test sets of that split. This way of measurement is based on the unnormalized difference, introduced in Section~\ref{sec:p_value}. We argue that using other methodologies, like those comparing sums of workload queries side by side, are statistically inaccurate, as they do not construct any hypothesis with an empirical and theoretical test metric, i.e. cannot provide significance in terms of confidence intervals. 

Note that the inference time of the ML models is not included in this first section, but we will analyze the results including the inference time in the following sections. The error bars indicate the statistical significance. Since a value of zero on the x-axis would correspond to an identical performance to PostgreSQL, any bar into the positive (right side) indicates that a method is faster than PostgreSQL. Conversely, bars into the negative (left side) of the x-axis indicate the opposite, i.e. the method is slower than PostgreSQL. Whenever the error bars cross over the zero position, the difference in performance is not statistically significant. In summary, whenever a bar is going to the positive side with an error bar that does not overlap the zero position, then that method outperforms PostgreSQL by a statistically significant margin.

% Analyze JOB
% Analyze Leave One Out Results - JOB
We start by looking at the results on JOB shown in Figure \ref{fig:results_job__no_inference}. On the \textit{leave one out split}, what is considered to be the easiest type of split, GenJoin is the only method that significantly outperforms PostgreSQL by 50 seconds taking 104 seconds on average, compared to 154 seconds for PostgreSQL. AutoSteer finishes the queries in 163 seconds, i.e. it is 9 seconds slower than PostgreSQL, while HybridQO requires 209 seconds, double the time of GenJoin.

% Analyze Random Results - JOB
The \textit{random split} shows an almost identical performance for all methods compared to the previous split without drastically changing the amount of time spent on planning and executing queries. GenJoin is 53 seconds faster than PostgreSQL on average, AutoSteer is 7 seconds slower than PostgreSQL, and HybridQO is 58 seconds slower. While some fluctuations are natural, this shows that the way the queries are distributed between leaving one out and random does not have an impact on the models' performances.

% Analyze Base Query Results - JOB
Finally, for the \textit{base query split}, there is less overlap between the seen training queries and the test queries. As for the previous splits, GenJoin demonstrates a stable lead over PostgreSQL being 49 seconds faster. However, HybridQO and AutoSteer have now switched ranks with HybridQO performing better than on other splits, albeit still 25 seconds slower than PostgreSQL. AutoSteer is most impacted by the different split method where queries are distributed to generate the adversarial case where most useful information between similar queries is hidden. AutoSteer now requires over 73 seconds more than PostgreSQL, although due to the large confidence intervals, AutoSteer is not statistically significantly slower, as there are large fluctuations in the three separate models trained.

% Analyze STACK
Now that we have analyzed the results for JOB, we take a look at the same results for STACK in Figure \ref{fig:results_stack__no_inference}. Compared to JOB, the queries in STACK are executed 50\% faster on average, which leaves less room to be optimized.

% Analyze Leave One Out Results - STACK
We will first analyze the \textit{leave one out split} for STACK, where both GenJoin and AutoSteer finish the queries faster than PostgreSQL. AutoSteer is 15 seconds and GenJoin 6 seconds faster than PostgreSQL, which took 81 seconds in total. HybridQO performs rather unstable with a massive confidence interval and takes 10 seconds longer than PostgreSQL.

% Analyze Random Results - STACK
On the \textit{random split} for STACK, we observe a nearly identical performance for GenJoin and AutoSteer, with both methods showing no change in their behavior in the different split types. Furthermore, HybridQO also outperforms PostgreSQL by 7 seconds. In summary,  for the random split all three methods outperform PostgreSQL by a statistically significant margin.

% Analyze Base Query Results - STACK
Finally, for the \textit{base query split}, GenJoin cements its stable mode of operation, 7 seconds faster than PostgreSQL. HybridQO is in second place, being 3 seconds faster. Lastly, AutoSteer's performance once again suffers on this split, even though it was the best method on the previous two splits for STACK. Even so, it still reaches a comparable time to PostgreSQL being just 2 seconds slower.

Over both JOB and STACK, we have observed that GenJoin is the only method that consistently outperforms PostgreSQL on all splits by a statistically significant amount of time. While AutoSteer finds better plans on two out of three splits on STACK compared to GenJoin, it never outperforms PostgreSQL on JOB. Similarly, HybridQO also displays a much more unstable performance that is much worse on JOB and barely better than GenJoin on one split of STACK. These results not only show the strength of our method in finding better plans but especially highlight an even more important ability: stability in different environments. However, across all evaluated LQOs there exist queries with degraded performance when analyzed individually.

\subsection{Measuring Query Processing Time}

In the following paragraphs, we outline the different measurements in evaluating an \textit{end-to-end execution time}, which includes: %(shown in Figure \ref{fig:end_to_end_execution_time})

\begin{enumerate}
    \item \textbf{Inference Time}: Any amount of time spent on other data processing that is not directly part of planning or executing a query (primarily for LQOs).
    \item \textbf{Planning Time}: The time investment by the RDBMS to analyze a query and optimize its query plans. Note, that LQOs do not skip the planning time, as the RDBMS optimizer applies further optimizations.
    \item \textbf{Execution Time}: How long the database engine takes to execute the physical plan and extract the result set.
\end{enumerate}

The primary goal of evaluating query optimizers based on the end-to-end execution time is to include all significant parts of the overall time spent in query execution while at the same time focusing only on the aspects where classical and LQO methods can have a direct impact. As such, network latency (i.e. in establishing a connection and sending the query to the RDBMS) is not included.

\begin{comment}
\begin{figure}[h!]
    \centering
    \includegraphics[width=\linewidth, keepaspectratio]{images/fig__end_to_end_execution_time.png}
    \caption{Break down of the end-to-end query execution time in an RDBMS for learned query optimizers. The end-to-end execution time consists of the inference, planning and execution time. The time spent on networking is omitted since it cannot be influenced by the LQO.}
    \label{fig:end_to_end_execution_time}
\end{figure}
\end{comment}

% JOB/STACK-Slow stuff is going to be shown here
\subsubsection{Inference Time}

With the introduction of LQO methods, there is now a need to also include inference time in the evaluation of the execution time, as learned methods can take a significant amount of time to produce their result\footnote{We focus on the amount of time required to process a \textit{previously unseen} query.} (e.g., encoding a query, using an ML model for inference or deciding what set of hints to pass to the RDBMS), before the RDBMS plans and executes a query. Inference time includes the prediction time of a model (e.g., neural network predictions), but also pre- and postprocessing steps (e.g., parsing the SQL statement, encoding a query into a vector representation for the model and turning it into an actionable hint set that can be prefixed to the original query) and additional EXPLAIN queries to fetch structural or statistical information from the classical optimizer (e.g., cardinality estimates for the encoding step). For a classical optimizer like the one used in PostgreSQL, this corresponds only to the planning time, as no additional processes are run. However, in the case of LQO methods, indicating this time as a separate entity allows us to compare the impact of the methods on the overall amount of time spent waiting for a query result. For this reason, we urge the community to focus not only on LQOs that produce good plans but also on methods that are efficient at inference.

With this new perspective, the goal of query optimization evaluations changed: While the execution time remains the primary metric to improve, it becomes equally important to find the plans, hint sets, or other LQO results in a reasonable amount of time. This begs the question: What is a \textit{reasonable} amount of time?

When considering inference time as a variable amount of time with a lower bound (e.g., because the encoding requires an EXPLAIN call that is on the order of 50 milliseconds), then the LQO should improve the execution time by at least the same amount to not become slower than the classical optimizer.\footnote{A different perspective is to say, that the amount of time spent categorized as inference time would not have existed if an RDBMS did not use an LQO.} %Depending on the context of one's workload, this lower bound implies that LQOs become increasingly suitable the longer the queries in a workload take to execute, and that small, efficient model architectures have an easier time realizing any performance gain. Unlike many other domains, this means that large language models in the current form are not the obvious choice of model architecture.

\subsubsection{Planning Time}

Arguably a smaller subset of the overall time spent executing queries is the planning time of the RDBMS. For LQOs that give hints or otherwise restrict the space of options, we could assume that less time is spent compared to the classical baseline. However, it has been shown that even in those cases, the planning time differs greatly and can grow by a factor of up to 5 depending on the method and workload \cite{lehmann2024your}. We want to stress that just because an LQO produces a plan for the RDBMS to execute, does not mean that the planning time can be skipped. %In fact, there remain many additional optimizations that the classical optimizer can do on top, especially in cases where the LQO only provides a small set of hints (e.g., no merge joins) on the optimization space. To the best of our knowledge, no LQO provides such a detailed physical plan to the engine, such that the internal optimization could be skipped.

\subsubsection{Execution Time}

Naturally, execution time is the primary optimization objective with the most impact for optimizers (classical and learned). We observe, in particular for the STACK workload, a high variance in execution times, which exacerbates the problem of achieving consistent measurements.

\subsection{Revisiting Results under Inference Time}
\label{sub:revisiting_with_inference}

Having discussed the importance of including inference time, we now revisit the results by measuring the end-to-end execution time of all methods against PostgreSQL. The goal is to assess if the LQOs not only find a plan that executes faster but also that the methods do so efficiently such that the overall execution time is reduced.

\begin{figure}[h!]
    \centering
    \includesvg[width=\linewidth, keepaspectratio]{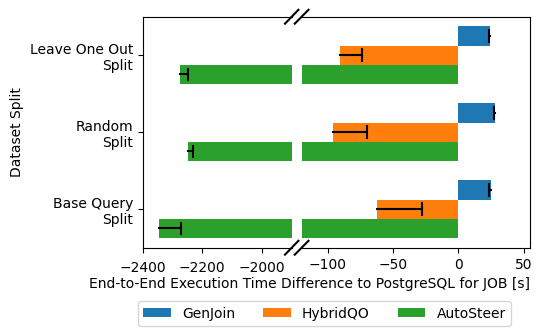}
    \caption{Results on the JOB using end-to-end execution times (incl. inference). Note the break on the x-axis.}
    \label{fig:results_job__with_inference}
\end{figure}

% Analyze JOB
We start with the Join Order Benchmark again, for which the results could be found in Figure \ref{fig:results_job__with_inference} now using the end-to-end execution time (i.e., including inference time).

On the \textit{leave one out split} GenJoin is still faster than PostgreSQL, but now only by 24 seconds. That means, half of the advantage vanishes in using GenJoin and finding an improved plan. Both HybridQO and AutoSteer are now significantly slower than PostgreSQL, with HybridQO needing an additional 36 seconds for finding its solutions. Finally the biggest jump is seen in AutoSteer, where the inference takes on the order of 37 minutes. To some extent, this is not surprising, as AutoSteer tries out a multitude of plans through EXPLAIN calls to find the hint set configurations, which lead to different plans than the PostgreSQL default plan. Since it is required to gather a subset of plans that differ to predict on an unseen query, this time is included in its inference time (and represents the dominant factor).

Looking at the \textit{random} and \textit{base query splits} reveals nothing new: GenJoin consistently loses about half its benefit in inference time (around 25 seconds), but can outperform PostgreSQL on all splits. HybridQO stays slower than PostgreSQL by an additional 37 seconds of inference time across both splits. AutoSteer again uses about 37 minutes for inference remaining in the last spot.

Interestingly, these results indicate that the way the queries are split into train and test sets has no impact on the amount of time spent producing a result, showing these times can be seen as a constant across a workload.

\begin{figure}[h!]
    \centering
    \includesvg[width=\linewidth, keepaspectratio]{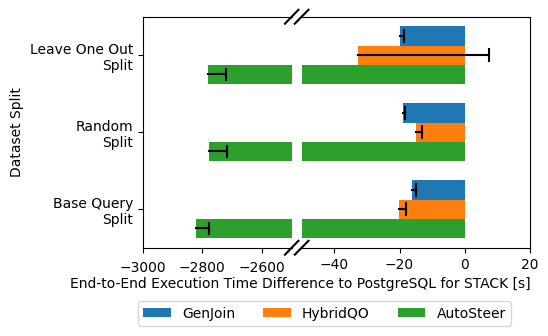}
    \caption{Results on STACK using end-to-end execution times (incl. inference). Note the break on the x-axis.}
    \label{fig:results_stack__with_inference}
\end{figure}

% Analyze STACK
Now we will analyze the results on STACK using Figure \ref{fig:results_stack__with_inference}. As we have discussed in Section \ref{sub:comparison_state_of_art}, the queries in the STACK workload leave less room for improvement in terms of absolute time spent. We can observe now that all methods, including GenJoin, are not able to beat PostgreSQL anymore on any type of split, with GenJoin and HybridQO reaching similar end-to-end execution times overall. While GenJoin's inference time is very consistent at around 24 seconds, HybridQO was able to find its result more quickly for STACK in 22 seconds, compared to 36 seconds for JOB. AutoSteer needs even more time for inference at 47 minutes.

\begin{comment}
\begin{figure}[h!]
    \centering
    \includegraphics[width=.9\linewidth, keepaspectratio]{images/fig__execution_time_pdf.png}
    \caption{Probability density function plot of the execution times of JOB and STACK, demonstrating that STACK contains a larger subset of queries that take less than one second.}
    \label{fig:execution_time_pdf}
\end{figure}
\end{comment}

We attribute this drastic change to the composition of queries in STACK, which includes about 50\% more queries than JOB that take less than a second. % In Figure \ref{fig:execution_time_pdf} we show a probability density function plot of the execution times of PostgreSQL on both workloads, where faster queries dominate STACK. 
The queries impose a harsh limit on the amount of time that can be used for inference and require that LQOs massively improve over the execution time of PostgreSQL. Due to the way current LQOs encode queries using EXPLAIN calls (which is typically the dominant factor for inference time) to gather auxiliary information from PostgreSQL, it remains an open question of how to encode queries and plans more efficiently.

\begin{figure}[h!]
    \centering
    \includesvg[width=0.88\linewidth, keepaspectratio]{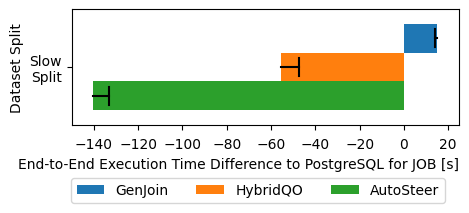}
    \caption{Results on JOB using end-to-end execution times (incl. inference) showing the \textit{slow split}.}
    \label{fig:results_job__with_inference_and_slow}
\end{figure}

\textit{Can a LQO outperform PostgreSQL on STACK?} To answer this question, we can first look at the \textit{slow split} with the 19 slowest queries from JOB, where the results are identical to the other splits (refer to Figure \ref{fig:results_job__with_inference}): GenJoin significantly outperforms PostgreSQL (see Figure \ref{fig:results_job__with_inference_and_slow}). However, both HybridQO and AutoSteer do not outperform PostgreSQL due to their large inference time.

\begin{figure}[h!]
    \centering
    \includesvg[width=0.88\linewidth, keepaspectratio]{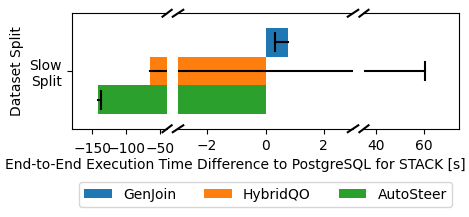}
    \caption{Results on STACK using end-to-end execution times (incl. inference) showing the \textit{slow split}. Note the two breaks on the x-axis to accommodate all orders of magnitude.}
    \label{fig:results_stack__with_inference_and_slow}
\end{figure}    

Figure \ref{fig:results_stack__with_inference_and_slow} illustrates the performance on the slow split of STACK, which contains the 19 slowest queries in its test set. We see now that GenJoin can overcome the constraints on the inference time and significantly outperform PostgreSQL. In particular, the test queries take long enough such that it becomes worthwhile to use GenJoin. HybridQO demonstrates a large variance in its end-to-end execution time due to instabilities in its predictions but takes over 60 seconds on average longer than PostgreSQL. Finally, AutoSteer spends over two minutes longer than PostgreSQL.

The experiments have shown that while both HybridQO and AutoSteer are able to find good plans, their inference time becomes a bottleneck to match or outperform PostgreSQL. Additionally, the different split types have shown large fluctuations in these methods' performances, with AutoSteer especially struggling on the adversarially constructed base query split. While GenJoin demonstrates a very stable performance across both workloads and various split types, it did not find the best plans on STACK. These results show the need for more efficient encoding approaches that do not require EXPLAIN calls, lowering the overhead cost of inference time for fast queries, posing the question of whether an LQO is even necessary for queries that are expected to finish in $< 1$ second.

%We attribute this drastic change to the already very quick queries that make up STACK which demonstrate clearly how inference time needs to be included to further progress learned query optimizers. In order to check our hypothesis, we have also performed experiments on the \textit{slow split}, originally introduced by the Balsa method \cite{Balsa_Yang2022} for JOB, and now extended to STACK by us. 

%In Figure \ref{fig:results_stack__with_inference_and_slow} we have added the results for GenJoin on all types of splits for STACK, including the \textit{slow split}. We can observe that by changing the test set to include the slowest queries, e.g., the queries with the most amount of time potentially optimized, GenJoin is able to significantly outperform PostgreSQL on that particular split. 

% This section gets the experiments that show the effect of chains of plans, and that the method is relatively stable and doesn't really degrade
\subsection{Ablation Study: Chain of Subplan Hints}
\label{sub:ablation_chain_of_plans}

With a method that generates a set of subplan hints given a set of subplan hints, the question arises if applying the model multiple times on its own output might yield even better results. This approach referred to as "chain-of-thought prompting" \cite{wei2022chain} in the space of large language models (LLMs) such as OpenAI's ChatGPT 4o model\footnote{\url{https://openai.com/index/learning-to-reason-with-llms/}}, has shown to be rather effective in guiding the LLM to better responses\footnote{E.g., the infamous "let's think step by step" prompt.}.
We have conducted an experiment where we let GenJoin start with a random set of subplan hints $H_0$ to predict the first iteration $H_1$ and, from then on, always use the previous output $H_{n-1}$ as the subplan hints input for $H_n$ (up to $N=25$).

The major drawback of the chain of subplan hints idea is that (a) it is not trivial to find the optimal number of iterations for each query and (b) with every additional iteration we spend more time making inference, which is already a limiting factor. However, the most costly aspect of GenJoin's inference time lies in the EXPLAIN queries to gather cardinality estimates from the RDBMS. To run additional iterations, however, the query encoding remains constant, and no additional EXPLAIN calls need to be executed, drastically speeding up any iterations after the first.

We have not observed a degradation in the execution time across 25 iterations, but rather minor fluctuations which could be explained by the natural swings in execution time. We also observed no significant improvement on the subplan hints generated after more than one iteration, showing that the additional time spent doing inference is not worth it for the overall end-to-end performance. Moreover, the results demonstrate the stability of the subplan hints generated by our method, confirming the validity of using randomly generated subplan hints over a pre-optimized set of hints, such as e.g., the set of hints contained in PostgreSQL's physical plan.

% ========================================================================================================

\section{Related Work}
\label{sec:related_work}

% Change the introductory setnence to include the mention of meta-lqos and environments, and then do shotgun citations for them.

In this section, we offer a short introduction to LQOs. We observe two distinct groups of methods that generate \textit{full plan-level hints} and \textit{query-level hints} (see Figure \ref{fig:genjoin_vs_others}), a number of \textit{meta-LQO} methods \cite{weng2024eraser, ammerlaan2021perfguard}, and \textit{environments} for ML-driven query optimization \cite{wang2024joingym, zhu2024pilotscope}. 

% TODO add full plan-level hints (Neo, Balsa, HybridQO, LEON, ...)
%First progress in the space of ML-based query optimization has been made through RL in methods like
The first ML-based query optimizers that apply RL are DQ \cite{DQ_krishnan2018learning}, ReJOIN \cite{ReJOIN_marcus2018deep, marcus2018rejoin1}, and FOOP \cite{heitz2019join}, predicting \textit{full plan-level hints}. To find the optimal join order, an exploration-exploitation strategy was applied using a reward cost model. 

One family of methods originated from Neo \cite{Neo_Marcus2019}, which was the first end-to-end method using a neural network to predict the latencies of a full query plan combined with a greedy bottom-up plan creation. Balsa \cite{Balsa_Yang2022} followed the overall structure of Neo, but introduced a variety of modifications to the training procedure, most prominently timeouts for query executions. 

Another family of LQOs started with RTOS \cite{RTOS_yu2020}, a method focusing on the sequence of two-way join operations disregarding join and scan types. Compared to Neo, RTOS uses a depth-first search for plan construction, also using an RL agent. LOGER \cite{chen2023loger} follows RTOS, extending the action space by including the types of joins. 

A new paradigm of changing the space of RL-based latency predictors was the emergence of learning-to-rank methods, such as Lero \cite{zhu2023lero} and LEON \cite{chen2023leon}. Lero generates a number of candidate plans by feeding alternate cardinality estimations into the PostgreSQL optimizer. A plan comparator model then predicts, given a pair of plans, which one to execute. LEON generates candidate plans through brute-forcing and pruning before the model ranks the candidate plans based on their estimated latency and uncertainty. LEON can be seen as the learning-to-rank continuation of Balsa.

Finally, HybridQO \cite{yu2022hybridqo}, as the name implies, combines cost and latency predictions. It generates candidate plans through hinting with the help of a Monte Carlo tree search where a model based on RTOS predicts their costs. The same type of network is then used to predict both the latency and uncertainty to inform the final model in choosing which plan to execute.

Now we will focus on the LQO methods that generate \textit{query-level hints}, emphasizing the main differences between GenJoin and other hints-based methods. In addition, we refer to \cite{thiessat2024hints} for a more detailed review of optimizer hinting.

Bao~\cite{Bao_Marcus2022} was the first method that does not produce complete query plans from scratch but uses \textit{hint sets} that prune the optimization space of a given plan. In total, Bao uses a set of 6 scan and join type hints (though they experimented with up to $2^6$ combinations and chose the best-performing subset). However, the hints are on a query level, rather than on a subplan or table level. %For example, Bao might disable any merge joins for the whole query, while GenJoin gives hints at a more granular level of two-way joins. 

COOOL~\cite{xu2023coool} uses pre-defined hint sets as a model input (contrary to other hint sets-based methods like Bao for which the hint set is the output of ML model) along with the query and plan encodings, and applies a learning-to-rank model to choose the best hint set according to the estimated execution time.

FASTgres~\cite{woltmann2023fastgres} uses hint set prediction like Bao (with the possibility of many more sets, essentially $2^6$ as a result of multi-class classification), primarily modifying the ML model, choosing a regression model instead of a bandit optimizer.

AutoSteer~\cite{anneser2023autosteer} expands upon the idea of Bao, where the space of query optimization rules is explored to find the best set of hints that globally enables or disables parts of the optimizer. While Bao uses 6 different scan and join hints, AutoSteer extends these to 20 hints for PostgreSQL (resulting in potential search space of $2^{20}$ hint sets), massively increasing the combinatorial space to be explored. AutoSteer further experiments on a variety of RDBMS where the number of hints (also called knobs) varies by an order of magnitude across systems. Instead of checking all combinations, AutoSteer iteratively checks if the hints make the optimizer deviate from its initial optimized plan and further merges these hints if they do. 

GenJoin considers a more limited amount of hints, in particular for the types of join, and applies them for each relation individually, rather than globally. GenJoin also does not need to explore which combinations of hints would make the optimizer deviate from its original plan, significantly reducing the amount of time spent optimizing an unseen query.

\begin{figure*}[h!t]
    \centering
    \includesvg[width=0.9\textwidth, keepaspectratio]{images/genjoin_pg_plan_change.drawio.svg}
    \caption{Illustration of the difference between the plans selected by GenJoin and PostgreSQL for Query 16b from the Join Order Benchmark. Differences at the beginning of the join tree have significant impact on the final three joins, since the Bitmap Index Scan loses the indexed ordering required by downstream nested loop joins that rely on index conditions.}
    \label{fig:genjoin_vs_pg}
\end{figure*}

\section{Discussion and Lessons Learned}

% "Things we exploited that work" / "Solved problem"
% - Symbiosis instead of Competition (exploited by GJ)
% Latest methods serve to reduce the space of potential solutions for PostgreSQL
% Effectiveness of including the built-in optimizer's strengths and years of research (colaboration over competition)
%      \item Accelerating an existing query optimizer through hints
%      \item Solving a part of the query optimization problem, but not replacing the optimizer
\textbf{Symbiosis instead of competition.} While the trend of working in tandem with the built-in optimizer is not completely new, designing the GenJoin method to only restrict its operations through hints and leaving a high degree of freedom in the hand of the built-in optimizer has proven to work well. This allows the LQO to focus on a more narrow problem, in turn reducing the inference time and improving the generalization ability.

% - Introducing additional proxy measurements (exploited by GJ)
%      \item Relative rankings instead of “just” execution time
% Found way to define a distance between the plans 
% It can serve both in the training data and as a distance metric for evaluation LQOs
% Highlight the properties, what is good about it (bounded, no unit, outlier robust, ...)
\textbf{A new distance metric for comparing query plans.} Not only has the T-test comparison solved the problem of stabilizing the range in which we compare plans in GenJoin's training data generation, but we could also show its effectiveness in evaluating LQO methods. Thanks to this, calculating the statistical significance has become straightforward. Moreover, this distance metric produces a bounded output that has no units, is robust to outliers, and allows for the interpretability of the p-values.

% - Explicit pruning vs 2-step architectures / Complexity Discussion from Fig1
\textbf{Tackling the complexity of the prediction space.} Analyzing the methods based on the categorization in Figure \ref{fig:genjoin_vs_others}, we identify three different classes of complexity for the space of potential predictions: full plan-level hint methods with $O(T^T)$\footnote{There are $T!$ different permutations for $T$ tables and for each permutation there are $C(T-1)$ possible binary trees, where $C()$ is the Catalan number. Simplifying, we end up with asymptotic complexity of $O(T^{T})$.}, query-level hint methods with $O(2^K)$ and GenJoin with $O(T^2)$ ($T$ represents the number of tables and $K$ the number of different hint sets). The first two classes require to process a potentially exponentially growing number of solutions. Because of this, they adopt a two-model approach, where one model prunes the search space and the other model finds the best solution in the pruned space (or vice versa).

GenJoin scales in the worst case polynomially with the number of tables, which does not require an explicit pruning step, allowing GenJoin to spend less time both for training and inference. In addition, having fewer components and trained parameters simplifies the overall architecture and leads to a more robust solution. Nevertheless, GenJoin follows the paradigm of a two-step approach explicitly by pruning the space of potential predictions during the training data generation, keeping only the training pairs for which the difference in execution time is statistically significant.

% - Discussing the instability of ANALYZE / Getting cardinality estimations from PG
% - - Leads to retraining the model after ANALYZE, since the model's performance is strongly correlated with the "DB state of ANALYZE"
\textbf{Cardinality estimations and \texttt{ANALYZE}.} It is evident that the way the internal statistics of PostgreSQL are sampled has a profound effect on the performance of the built-in optimizer. LQOs are equally affected by changes in the internal statistics, for example, from explicit \texttt{ANALYZE} calls or through incidental updates via \texttt{AUTOVACUUM}, as their performance is strongly correlated with the internal state of PostgreSQL's statistics. An interesting consequence of this realization is that reproducibility is further challenged not only by the available hardware, the database configuration, and the random seed during the training of an ML model but also by the sampled values of PostgreSQL's internal statistics.

\textbf{The nature of GenJoin's advantage}. In the majority of the cases where GenJoin significantly outperforms PostgreSQL, there is a common pattern of query plan tree change. We can demonstrate it using as an example query 16b from JOB. This query behaves stably among the splits, resulting in roughly 13.2 seconds for PostgreSQL and 4.3 seconds for GenJoin. We have depicted the key differences of the query plan trees between PostgreSQL and GenJoin in Figure \ref{fig:genjoin_vs_pg}. The only deviating part is a first action at a left deep tree, where PostgreSQL chooses to perform $NL(k, mk)$ and GenJoin prefers starting with $HJ(mk, k)$. Such a tiny, at first glance, change led to drastic consequences, since (1) for the nested loop join PostgreSQL used a fast BitMap Index Scan over $mk$, while GenJoin had to switch and run a longer Seq Scan (as the join type defines the scan type as discussed in Section~\ref{sec:intro}); (2) as a result, for PostgreSQL the $mk$ lost the original indexing~\cite{bitmapindexscan}, and could not be used efficiently for further Index Scans by $ci$, $n$ and $an$ - for GenJoin it is not an issue. This way, we see a clear example of how classical greedy optimizers might win in subplan performance (local optimum), while the generative approach would outperform on the plan level (global optimum).

\textbf{Lessons from competitor performance}. Both HybridQO and AutoSteer consistently predict changes from the default PostgreSQL plan, even if the plan chosen by PostgreSQL is optimal. In fact, between $5$ and $13\%$ of queries (depending on the dataset split) are best executed without changing the default plan. These queries seem to drive the observed performance degradation on the base query split for AutoSteer. Furthermore, HybridQO always predicts a single \texttt{Leading(t1 t2)} hint specifying the first two-way join. Unlike AutoSteer, HybridQO can propose a hint that matches PostgreSQL's default plan, which happens in approximately $21\%$ of cases.

\section{Conclusion and Future Work}

We have introduced GenJoin, the first generative LQO that consistently outperforms PostgreSQL in terms of execution time. GenJoin is able to compete with and outperform current state-of-the-art methods, in particular in terms of the amount of time spent for inference. We have demonstrated that GenJoin behaves very stably across a variety of workloads and train/test split characteristics.

%\section{Future Work}

The experimental results have highlighted a need for encoding strategies that are more efficient at inference. We plan to investigate how GenJoin, or LQOs in general, can produce query and plan encodings that conserve a similar level of expressiveness without the need to call the RDBMS for cardinality estimations or other auxiliary information through EXPLAIN queries. In addition, there remain many situations that are badly captured by current encodings, such as tables appearing multiple times in the same query through multiple aliases, multiple attributes on which to join the same two tables, or performing self-referential joins.

%Additionally, since the subplan hints used in GenJoin are also supported by such well-spread RDBMS like Oracle and SQL Server, we are considering experimenting with those systems as one of the follow up steps.

%Additionally, we will explore how data drift affects the performance of GenJoin, in particular when the sampled internal statistics (e.g., through \texttt{AUTOVACUUM} or \texttt{ANALYZE}) have changed.

\begin{comment}
We further see the potential in GenJoin to replace PostgreSQL as its reference system and instead use a weaker model (e.g., a previous generation of trained models). This idea has already been explored as a trend in superalignment~\cite{burns2024weaktostrong}, but could be applied here as well.
\end{comment}
We acknowledge the recent trend in bringing LQOs into industrial applications \cite{anneser2023autosteer, zhu2024pilotscope, ammerlaan2021perfguard} through the use of meta-LQO methods \cite{weng2024eraser} and database environments \cite{wang2024joingym} for faster iteration and reduced performance regression. With the novel architecture of GenJoin and its training data generation, we will explore how to further apply it in more challenging settings, as well as in other RDBMS that support subplan hints (e.g., Oracle or SQL Server).

\begin{acks}
The project received funding from the Swiss National Science Foundation under grant number 192105. This work was also partially supported by DataGEMS, funded by the European Union's Horizon Europe Research and Innovation Programme, under grant agreement No 101188416.
\end{acks}

\vfill\null
\bibliographystyle{ACM-Reference-Format}
\bibliography{main}


\begin{thebibliography}{52}

%%% the Web address.

\ifx \showCODEN    \undefined \def \showCODEN     #1{\unskip}     \fi
\ifx \showISBNx    \undefined \def \showISBNx     #1{\unskip}     \fi
\ifx \showISBNxiii \undefined \def \showISBNxiii  #1{\unskip}     \fi
\ifx \showISSN     \undefined \def \showISSN      #1{\unskip}     \fi
\ifx \showLCCN     \undefined \def \showLCCN      #1{\unskip}     \fi
\ifx \shownote     \undefined \def \shownote      #1{#1}          \fi
\ifx \showarticletitle \undefined \def \showarticletitle #1{#1}   \fi
\ifx \showURL      \undefined \def \showURL       {\relax}        \fi
% invisible to TeX
\providecommand\bibfield[2]{#2}
\providecommand\bibinfo[2]{#2}
\providecommand\natexlab[1]{#1}
\providecommand\showeprint[2][]{arXiv:#2}

\bibitem[Ammerlaan et~al\mbox{.}(2021)]%
        {ammerlaan2021perfguard}
\bibfield{author}{\bibinfo{person}{Remmelt Ammerlaan}, \bibinfo{person}{Gilbert Antonius}, \bibinfo{person}{Marc Friedman}, \bibinfo{person}{HM~Sajjad Hossain}, \bibinfo{person}{Alekh Jindal}, \bibinfo{person}{Peter Orenberg}, \bibinfo{person}{Hiren Patel}, \bibinfo{person}{Shi Qiao}, \bibinfo{person}{Vijay Ramani}, \bibinfo{person}{Lucas Rosenblatt}, {et~al\mbox{.}}} \bibinfo{year}{2021}\natexlab{}.
\newblock \showarticletitle{PerfGuard: deploying ML-for-systems without performance regressions, almost!}
\newblock \bibinfo{journal}{\emph{Proceedings of the VLDB Endowment}} \bibinfo{volume}{14}, \bibinfo{number}{13} (\bibinfo{year}{2021}), \bibinfo{pages}{3362--3375}.
\newblock


\bibitem[Anneser et~al\mbox{.}(2023)]%
        {anneser2023autosteer}
\bibfield{author}{\bibinfo{person}{Christoph Anneser}, \bibinfo{person}{Nesime Tatbul}, \bibinfo{person}{David Cohen}, \bibinfo{person}{Zhenggang Xu}, \bibinfo{person}{Prithviraj Pandian}, \bibinfo{person}{Nikolay Laptev}, {and} \bibinfo{person}{Ryan Marcus}.} \bibinfo{year}{2023}\natexlab{}.
\newblock \showarticletitle{Autosteer: Learned query optimization for any sql database}.
\newblock \bibinfo{journal}{\emph{Proceedings of the VLDB Endowment}} \bibinfo{volume}{16}, \bibinfo{number}{12} (\bibinfo{year}{2023}), \bibinfo{pages}{3515--3527}.
\newblock


\bibitem[Avnur and Hellerstein(2000)]%
        {hellerstein2000dynamic}
\bibfield{author}{\bibinfo{person}{Ron Avnur} {and} \bibinfo{person}{Joseph~M. Hellerstein}.} \bibinfo{year}{2000}\natexlab{}.
\newblock \showarticletitle{Eddies: Continuously Adaptive Query Processing}.
\newblock \bibinfo{journal}{\emph{SIGMOD Rec.}} \bibinfo{volume}{29}, \bibinfo{number}{2} (\bibinfo{date}{may} \bibinfo{year}{2000}), \bibinfo{pages}{261–272}.
\newblock
\showISSN{0163-5808}
\href{https://doi.org/10.1145/335191.335420}{doi:\nolinkurl{10.1145/335191.335420}}


\bibitem[Bishop(2006)]%
        {bishop2006generalization}
\bibfield{author}{\bibinfo{person}{Christopher~M. Bishop}.} \bibinfo{year}{2006}\natexlab{}.
\newblock \bibinfo{booktitle}{\emph{Pattern Recognition and Machine Learning (Information Science and Statistics)}}.
\newblock \bibinfo{publisher}{Springer-Verlag}, \bibinfo{address}{Berlin, Heidelberg}.
\newblock
\showISBNx{0387310738}


\bibitem[Chen et~al\mbox{.}(2023b)]%
        {chen2023loger}
\bibfield{author}{\bibinfo{person}{Tianyi Chen}, \bibinfo{person}{Jun Gao}, \bibinfo{person}{Hedui Chen}, {and} \bibinfo{person}{Yaofeng Tu}.} \bibinfo{year}{2023}\natexlab{b}.
\newblock \showarticletitle{LOGER: A Learned Optimizer Towards Generating Efficient and Robust Query Execution Plans}.
\newblock \bibinfo{journal}{\emph{Proceedings of the VLDB Endowment}} \bibinfo{volume}{16}, \bibinfo{number}{7} (\bibinfo{year}{2023}), \bibinfo{pages}{1777--1789}.
\newblock


\bibitem[Chen et~al\mbox{.}(2023a)]%
        {chen2023leon}
\bibfield{author}{\bibinfo{person}{Xu Chen}, \bibinfo{person}{Haitian Chen}, \bibinfo{person}{Zibo Liang}, \bibinfo{person}{Shuncheng Liu}, \bibinfo{person}{Jinghong Wang}, \bibinfo{person}{Kai Zeng}, \bibinfo{person}{Han Su}, {and} \bibinfo{person}{Kai Zheng}.} \bibinfo{year}{2023}\natexlab{a}.
\newblock \showarticletitle{LEON: A New Framework for ML-Aided Query Optimization.}
\newblock \bibinfo{journal}{\emph{Proc. VLDB Endow.}} \bibinfo{volume}{16}, \bibinfo{number}{9} (\bibinfo{year}{2023}), \bibinfo{pages}{2261--2273}.
\newblock


\bibitem[Cormen et~al\mbox{.}(2009)]%
        {cormen2009algorithms}
\bibfield{author}{\bibinfo{person}{Thomas~H. Cormen}, \bibinfo{person}{Charles~E. Leiserson}, \bibinfo{person}{Ronald~L. Rivest}, {and} \bibinfo{person}{Clifford Stein}.} \bibinfo{year}{2009}\natexlab{}.
\newblock \bibinfo{booktitle}{\emph{Introduction to Algorithms, Third Edition} (\bibinfo{edition}{3rd} ed.)}.
\newblock \bibinfo{publisher}{The MIT Press}.
\newblock
\showISBNx{0262033844}


\bibitem[David M.~Blei and McAuliffe(2017)]%
        {blei2017vi}
\bibfield{author}{\bibinfo{person}{Alp~Kucukelbir David M.~Blei} {and} \bibinfo{person}{Jon~D. McAuliffe}.} \bibinfo{year}{2017}\natexlab{}.
\newblock \showarticletitle{Variational Inference: A Review for Statisticians}.
\newblock \bibinfo{journal}{\emph{J. Amer. Statist. Assoc.}} \bibinfo{volume}{112}, \bibinfo{number}{518} (\bibinfo{year}{2017}), \bibinfo{pages}{859--877}.
\newblock
\href{https://doi.org/10.1080/01621459.2017.1285773}{doi:\nolinkurl{10.1080/01621459.2017.1285773}}


\bibitem[Ding et~al\mbox{.}(2024)]%
        {ding2024lqos}
\bibfield{author}{\bibinfo{person}{Bolin Ding}, \bibinfo{person}{Rong Zhu}, {and} \bibinfo{person}{Jingren Zhou}.} \bibinfo{year}{2024}\natexlab{}.
\newblock \showarticletitle{Learned Query Optimizers}.
\newblock \bibinfo{journal}{\emph{Foundations and Trends® in Databases}} \bibinfo{volume}{13}, \bibinfo{number}{4} (\bibinfo{year}{2024}), \bibinfo{pages}{250--310}.
\newblock
\showISSN{1931-7883}
\href{https://doi.org/10.1561/1900000082}{doi:\nolinkurl{10.1561/1900000082}}


\bibitem[Fu(2006)]%
        {fu2006reparametrization}
\bibfield{author}{\bibinfo{person}{Michael~C. Fu}.} \bibinfo{year}{2006}\natexlab{}.
\newblock \showarticletitle{Chapter 19 Gradient Estimation}.
\newblock In \bibinfo{booktitle}{\emph{Simulation}}, \bibfield{editor}{\bibinfo{person}{Shane~G. Henderson} {and} \bibinfo{person}{Barry~L. Nelson}} (Eds.). \bibinfo{series}{Handbooks in Operations Research and Management Science}, Vol.~\bibinfo{volume}{13}. \bibinfo{publisher}{Elsevier}, \bibinfo{pages}{575--616}.
\newblock
\showISSN{0927-0507}
\href{https://doi.org/10.1016/S0927-0507(06)13019-4}{doi:\nolinkurl{10.1016/S0927-0507(06)13019-4}}


\bibitem[Graefe(1995)]%
        {graefe1995cascades}
\bibfield{author}{\bibinfo{person}{Goetz Graefe}.} \bibinfo{year}{1995}\natexlab{}.
\newblock \showarticletitle{The Cascades Framework for Query Optimization.}
\newblock \bibinfo{journal}{\emph{IEEE Data(base) Engineering Bulletin}}  \bibinfo{volume}{18} (\bibinfo{year}{1995}), \bibinfo{pages}{19--29}.
\newblock
\urldef\tempurl%
\url{https://api.semanticscholar.org/CorpusID:260706023}
\showURL{%
\tempurl}


\bibitem[Heitz and Stockinger(2019)]%
        {heitz2019join}
\bibfield{author}{\bibinfo{person}{Jonas Heitz} {and} \bibinfo{person}{Kurt Stockinger}.} \bibinfo{year}{2019}\natexlab{}.
\newblock \showarticletitle{Join query optimization with deep reinforcement learning algorithms}.
\newblock \bibinfo{journal}{\emph{arXiv preprint arXiv:1911.11689}} (\bibinfo{year}{2019}).
\newblock


\bibitem[Hilprecht and Binnig(2022)]%
        {hilprecht2022zero}
\bibfield{author}{\bibinfo{person}{Benjamin Hilprecht} {and} \bibinfo{person}{Carsten Binnig}.} \bibinfo{year}{2022}\natexlab{}.
\newblock \showarticletitle{Zero-shot cost models for out-of-the-box learned cost prediction}.
\newblock \bibinfo{journal}{\emph{arXiv preprint arXiv:2201.00561}} (\bibinfo{year}{2022}).
\newblock


\bibitem[Hilprecht et~al\mbox{.}(2019)]%
        {hilprecht2019deepdb}
\bibfield{author}{\bibinfo{person}{Benjamin Hilprecht}, \bibinfo{person}{Andreas Schmidt}, \bibinfo{person}{Moritz Kulessa}, \bibinfo{person}{Alejandro Molina}, \bibinfo{person}{Kristian Kersting}, {and} \bibinfo{person}{Carsten Binnig}.} \bibinfo{year}{2019}\natexlab{}.
\newblock \showarticletitle{Deepdb: Learn from data, not from queries!}
\newblock \bibinfo{journal}{\emph{arXiv preprint arXiv:1909.00607}} (\bibinfo{year}{2019}).
\newblock


\bibitem[Kendall and Stuart(1973)]%
        {kendall1973statistics}
\bibfield{author}{\bibinfo{person}{M.G. Kendall} {and} \bibinfo{person}{A. Stuart}.} \bibinfo{year}{1973}\natexlab{}.
\newblock \bibinfo{booktitle}{\emph{The Advanced Theory of Statistics. Vol. 2: Inference and: Relationsship}}.
\newblock \bibinfo{publisher}{Griffin}.
\newblock
\urldef\tempurl%
\url{https://books.google.ch/books?id=elabQwAACAAJ}
\showURL{%
\tempurl}


\bibitem[Kipf et~al\mbox{.}(2018)]%
        {MSCN_kipf2018learned}
\bibfield{author}{\bibinfo{person}{Andreas Kipf}, \bibinfo{person}{Thomas Kipf}, \bibinfo{person}{Bernhard Radke}, \bibinfo{person}{Viktor Leis}, \bibinfo{person}{Peter Boncz}, {and} \bibinfo{person}{Alfons Kemper}.} \bibinfo{year}{2018}\natexlab{}.
\newblock \showarticletitle{Learned cardinalities: Estimating correlated joins with deep learning}.
\newblock \bibinfo{journal}{\emph{arXiv preprint arXiv:1809.00677}} (\bibinfo{year}{2018}).
\newblock


\bibitem[Krishnan et~al\mbox{.}(2018)]%
        {DQ_krishnan2018learning}
\bibfield{author}{\bibinfo{person}{Sanjay Krishnan}, \bibinfo{person}{Zongheng Yang}, \bibinfo{person}{Ken Goldberg}, \bibinfo{person}{Joseph Hellerstein}, {and} \bibinfo{person}{Ion Stoica}.} \bibinfo{year}{2018}\natexlab{}.
\newblock \showarticletitle{Learning to optimize join queries with deep reinforcement learning}.
\newblock \bibinfo{journal}{\emph{arXiv preprint arXiv:1808.03196}} (\bibinfo{year}{2018}).
\newblock


\bibitem[Lane(2005)]%
        {bitmapindexscan}
\bibfield{author}{\bibinfo{person}{Tom Lane}.} \bibinfo{year}{2005}\natexlab{}.
\newblock \bibinfo{title}{{Re: Bitmap indexes etc.}}
\newblock \bibinfo{howpublished}{\url{https://www.postgresql.org/message-id/12553.1135634231@sss.pgh.pa.us}}.
\newblock
\newblock
\shownote{[Online; accessed April, 2025]}.


\bibitem[Lehmann et~al\mbox{.}(2024)]%
        {lehmann2024your}
\bibfield{author}{\bibinfo{person}{Claude Lehmann}, \bibinfo{person}{Pavel Sulimov}, {and} \bibinfo{person}{Kurt Stockinger}.} \bibinfo{year}{2024}\natexlab{}.
\newblock \showarticletitle{Is Your Learned Query Optimizer Behaving As You Expect? A Machine Learning Perspective}.
\newblock \bibinfo{journal}{\emph{Proceedings of the VLDB Endowment}} \bibinfo{volume}{17}, \bibinfo{number}{7} (\bibinfo{year}{2024}), \bibinfo{pages}{1565--1577}.
\newblock
\urldef\tempurl%
\url{https://dl.acm.org/doi/10.14778/3654621.3654625}
\showURL{%
\tempurl}


\bibitem[Leis et~al\mbox{.}(2015)]%
        {JOB_Leis2015}
\bibfield{author}{\bibinfo{person}{Viktor Leis}, \bibinfo{person}{Andrey Gubichev}, \bibinfo{person}{Atanas Mirchev}, \bibinfo{person}{Peter Boncz}, \bibinfo{person}{Alfons Kemper}, {and} \bibinfo{person}{Thomas Neumann}.} \bibinfo{year}{2015}\natexlab{}.
\newblock \showarticletitle{How good are query optimizers, really?}
\newblock \bibinfo{journal}{\emph{Proceedings of the VLDB Endowment}}  \bibinfo{volume}{9} (\bibinfo{date}{11} \bibinfo{year}{2015}), \bibinfo{pages}{204--215}.
\newblock
Issue 3.
\showISSN{21508097}
\href{https://doi.org/10.14778/2850583.2850594}{doi:\nolinkurl{10.14778/2850583.2850594}}


\bibitem[Lemons and Langevin(2002)]%
        {lemons2002sumofnormals}
\bibfield{author}{\bibinfo{person}{D.S. Lemons} {and} \bibinfo{person}{P. Langevin}.} \bibinfo{year}{2002}\natexlab{}.
\newblock \bibinfo{booktitle}{\emph{An Introduction to Stochastic Processes in Physics}}.
\newblock \bibinfo{publisher}{Johns Hopkins University Press}.
\newblock
\showISBNx{9780801868672}
\showLCCN{2001046459}
\urldef\tempurl%
\url{https://books.google.ch/books?id=Uw6YDkd_CXcC}
\showURL{%
\tempurl}


\bibitem[Li et~al\mbox{.}(2021)]%
        {li2021cardinality}
\bibfield{author}{\bibinfo{person}{Beibin Li}, \bibinfo{person}{Yao Lu}, \bibinfo{person}{Chi Wang}, {and} \bibinfo{person}{Srikanth Kandula}.} \bibinfo{year}{2021}\natexlab{}.
\newblock \showarticletitle{Cardinality Estimation: Is Machine Learning a Silver Bullet?}. In \bibinfo{booktitle}{\emph{AIDB}}.
\newblock
\urldef\tempurl%
\url{https://www.microsoft.com/en-us/research/publication/cardinality-estimation-is-machine-learning-a-silver-bullet/}
\showURL{%
\tempurl}


\bibitem[Li(2023)]%
        {shengbo2023dprl}
\bibfield{author}{\bibinfo{person}{Shengbo Li}.} \bibinfo{year}{2023}\natexlab{}.
\newblock \bibinfo{booktitle}{\emph{Reinforcement Learning for Sequential Decision and Optimal Control}}.
\newblock \bibinfo{publisher}{Springer}.
\newblock
\showISBNx{978-981-19-7783-1}
\href{https://doi.org/10.1007/978-981-19-7784-8}{doi:\nolinkurl{10.1007/978-981-19-7784-8}}


\bibitem[Liu et~al\mbox{.}(2021)]%
        {liu2021fauce}
\bibfield{author}{\bibinfo{person}{Jie Liu}, \bibinfo{person}{Wenqian Dong}, \bibinfo{person}{Qingqing Zhou}, {and} \bibinfo{person}{Dong Li}.} \bibinfo{year}{2021}\natexlab{}.
\newblock \showarticletitle{Fauce: fast and accurate deep ensembles with uncertainty for cardinality estimation}.
\newblock \bibinfo{journal}{\emph{Proceedings of the VLDB Endowment}} \bibinfo{volume}{14}, \bibinfo{number}{11} (\bibinfo{year}{2021}), \bibinfo{pages}{1950--1963}.
\newblock


\bibitem[Marcus(2023)]%
        {marcus2023learned}
\bibfield{author}{\bibinfo{person}{Ryan Marcus}.} \bibinfo{year}{2023}\natexlab{}.
\newblock \showarticletitle{Learned Query Superoptimization}.
\newblock \bibinfo{journal}{\emph{arXiv preprint arXiv:2303.15308}} (\bibinfo{year}{2023}).
\newblock


\bibitem[Marcus et~al\mbox{.}(2022)]%
        {Bao_Marcus2022}
\bibfield{author}{\bibinfo{person}{Ryan Marcus}, \bibinfo{person}{Parimarjan Negi}, \bibinfo{person}{Hongzi Mao}, \bibinfo{person}{Nesime Tatbul}, \bibinfo{person}{Mohammad Alizadeh}, {and} \bibinfo{person}{Tim Kraska}.} \bibinfo{year}{2022}\natexlab{}.
\newblock \showarticletitle{Bao: Making Learned Query Optimization Practical}.
\newblock \bibinfo{journal}{\emph{ACM SIGMOD Record}}  \bibinfo{volume}{51} (\bibinfo{date}{6} \bibinfo{year}{2022}), \bibinfo{pages}{6--13}.
\newblock
Issue 1.
\showISBNx{9781450383431}
\showISSN{01635808}
\href{https://doi.org/10.1145/3542700.3542703}{doi:\nolinkurl{10.1145/3542700.3542703}}


\bibitem[Marcus et~al\mbox{.}(2019)]%
        {Neo_Marcus2019}
\bibfield{author}{\bibinfo{person}{Ryan Marcus}, \bibinfo{person}{Parimarjan Negi}, \bibinfo{person}{Hongzi Mao}, \bibinfo{person}{Chi Zhang}, \bibinfo{person}{Mohammad Alizadeh}, \bibinfo{person}{Tim Kraska}, \bibinfo{person}{Olga Papaemmanouil}, {and} \bibinfo{person}{Nesime Tatbul}.} \bibinfo{year}{2019}\natexlab{}.
\newblock \showarticletitle{Neo: A Learned Query Optimizer}.
\newblock \bibinfo{journal}{\emph{Proceedings of the VLDB Endowment}}  \bibinfo{volume}{12} (\bibinfo{date}{4} \bibinfo{year}{2019}), \bibinfo{pages}{1705--1718}.
\newblock
Issue 11.
\href{https://doi.org/10.14778/3342263.3342644}{doi:\nolinkurl{10.14778/3342263.3342644}}


\bibitem[Marcus and Papaemmanouil(2018a)]%
        {ReJOIN_marcus2018deep}
\bibfield{author}{\bibinfo{person}{Ryan Marcus} {and} \bibinfo{person}{Olga Papaemmanouil}.} \bibinfo{year}{2018}\natexlab{a}.
\newblock \showarticletitle{Deep reinforcement learning for join order enumeration}. In \bibinfo{booktitle}{\emph{Proceedings of the First International Workshop on Exploiting Artificial Intelligence Techniques for Data Management}}. \bibinfo{pages}{1--4}.
\newblock


\bibitem[Marcus and Papaemmanouil(2018b)]%
        {marcus2018rejoin1}
\bibfield{author}{\bibinfo{person}{Ryan Marcus} {and} \bibinfo{person}{Olga Papaemmanouil}.} \bibinfo{year}{2018}\natexlab{b}.
\newblock \showarticletitle{Towards a Hands-Free Query Optimizer through Deep Learning}.
\newblock  (\bibinfo{year}{2018}).
\newblock
\showeprint[arxiv]{1809.10212}~[cs.DB]
\urldef\tempurl%
\url{https://arxiv.org/abs/1809.10212}
\showURL{%
\tempurl}


\bibitem[Panigrahy(2008)]%
        {[panigraphy2008kdtree]}
\bibfield{author}{\bibinfo{person}{Rina Panigrahy}.} \bibinfo{year}{2008}\natexlab{}.
\newblock \showarticletitle{An improved algorithm finding nearest neighbor using Kd-trees}. In \bibinfo{booktitle}{\emph{Proceedings of the 8th Latin American Conference on Theoretical Informatics}} (B\'{u}zios, Brazil) \emph{(\bibinfo{series}{LATIN'08})}. \bibinfo{publisher}{Springer-Verlag}, \bibinfo{address}{Berlin, Heidelberg}, \bibinfo{pages}{387–398}.
\newblock
\showISBNx{3540787720}


\bibitem[Reiner and Grossniklaus(2023)]%
        {reiner2023sample}
\bibfield{author}{\bibinfo{person}{Silvan Reiner} {and} \bibinfo{person}{Michael Grossniklaus}.} \bibinfo{year}{2023}\natexlab{}.
\newblock \showarticletitle{Sample-Efficient Cardinality Estimation Using Geometric Deep Learning}.
\newblock \bibinfo{journal}{\emph{Proceedings of the VLDB Endowment}} \bibinfo{volume}{17}, \bibinfo{number}{4} (\bibinfo{year}{2023}), \bibinfo{pages}{740--752}.
\newblock


\bibitem[Sohn et~al\mbox{.}(2015)]%
        {sohn2015learning_CVAE}
\bibfield{author}{\bibinfo{person}{Kihyuk Sohn}, \bibinfo{person}{Honglak Lee}, {and} \bibinfo{person}{Xinchen Yan}.} \bibinfo{year}{2015}\natexlab{}.
\newblock \showarticletitle{Learning structured output representation using deep conditional generative models}.
\newblock \bibinfo{journal}{\emph{Advances in neural information processing systems}}  \bibinfo{volume}{28} (\bibinfo{year}{2015}).
\newblock


\bibitem[Student(1908)]%
        {student1908ttest}
\bibfield{author}{\bibinfo{person}{Student}.} \bibinfo{year}{1908}\natexlab{}.
\newblock \showarticletitle{The probable error of a mean}.
\newblock \bibinfo{journal}{\emph{Biometrika}} (\bibinfo{year}{1908}), \bibinfo{pages}{1--25}.
\newblock


\bibitem[Sutton and Barto(2018)]%
        {sutton2018rl}
\bibfield{author}{\bibinfo{person}{Richard~S Sutton} {and} \bibinfo{person}{Andrew~G Barto}.} \bibinfo{year}{2018}\natexlab{}.
\newblock \bibinfo{booktitle}{\emph{Reinforcement learning: An introduction}}.
\newblock \bibinfo{publisher}{MIT press}.
\newblock


\bibitem[Thiessat et~al\mbox{.}(2024)]%
        {thiessat2024hints}
\bibfield{author}{\bibinfo{person}{Jerome Thiessat}, \bibinfo{person}{Dirk Habich}, {and} \bibinfo{person}{Wolfgang Lehner}.} \bibinfo{year}{2024}\natexlab{}.
\newblock \showarticletitle{Steering the PostgreSQL query optimizer using hinting: State-Of-The-Art and open challenges}.
\newblock  (\bibinfo{date}{06} \bibinfo{year}{2024}).
\newblock


\bibitem[Thorndike(1953)]%
        {thorndike1953elbow}
\bibfield{author}{\bibinfo{person}{Robert~L. Thorndike}.} \bibinfo{year}{1953}\natexlab{}.
\newblock \showarticletitle{Who belongs in the family?}
\newblock \bibinfo{journal}{\emph{Psychometrika}}  \bibinfo{volume}{18} (\bibinfo{year}{1953}), \bibinfo{pages}{267--276}.
\newblock
\urldef\tempurl%
\url{https://api.semanticscholar.org/CorpusID:120467216}
\showURL{%
\tempurl}


\bibitem[Wang and Chen(1996)]%
        {wang1996complexity}
\bibfield{author}{\bibinfo{person}{Chihping Wang} {and} \bibinfo{person}{Ming-Syan Chen}.} \bibinfo{year}{1996}\natexlab{}.
\newblock \showarticletitle{On the complexity of distributed query optimization}.
\newblock \bibinfo{journal}{\emph{IEEE Transactions on Knowledge and Data Engineering}} \bibinfo{volume}{8}, \bibinfo{number}{4} (\bibinfo{year}{1996}), \bibinfo{pages}{650--662}.
\newblock


\bibitem[Wang et~al\mbox{.}(2024)]%
        {wang2024joingym}
\bibfield{author}{\bibinfo{person}{Junxiong Wang}, \bibinfo{person}{Kaiwen Wang}, \bibinfo{person}{Yueying Li}, \bibinfo{person}{Nathan Kallus}, \bibinfo{person}{Immanuel Trummer}, {and} \bibinfo{person}{Wen Sun}.} \bibinfo{year}{2024}\natexlab{}.
\newblock \bibinfo{title}{JoinGym: An Efficient Query Optimization Environment for Reinforcement Learning}.
\newblock
\urldef\tempurl%
\url{https://openreview.net/forum?id=aAEBTnTGo3}
\showURL{%
\tempurl}


\bibitem[Wei et~al\mbox{.}(2022)]%
        {wei2022chain}
\bibfield{author}{\bibinfo{person}{Jason Wei}, \bibinfo{person}{Xuezhi Wang}, \bibinfo{person}{Dale Schuurmans}, \bibinfo{person}{Maarten Bosma}, \bibinfo{person}{Fei Xia}, \bibinfo{person}{Ed Chi}, \bibinfo{person}{Quoc~V Le}, \bibinfo{person}{Denny Zhou}, {et~al\mbox{.}}} \bibinfo{year}{2022}\natexlab{}.
\newblock \showarticletitle{Chain-of-thought prompting elicits reasoning in large language models}.
\newblock \bibinfo{journal}{\emph{Advances in neural information processing systems}}  \bibinfo{volume}{35} (\bibinfo{year}{2022}), \bibinfo{pages}{24824--24837}.
\newblock


\bibitem[Weng et~al\mbox{.}(2024)]%
        {weng2024eraser}
\bibfield{author}{\bibinfo{person}{Lianggui Weng}, \bibinfo{person}{Rong Zhu}, \bibinfo{person}{Di Wu}, \bibinfo{person}{Bolin Ding}, \bibinfo{person}{Bolong Zheng}, {and} \bibinfo{person}{Jingren Zhou}.} \bibinfo{year}{2024}\natexlab{}.
\newblock \showarticletitle{Eraser: Eliminating Performance Regression on Learned Query Optimizer}.
\newblock \bibinfo{journal}{\emph{Proc. VLDB Endow.}} \bibinfo{volume}{17}, \bibinfo{number}{5} (\bibinfo{date}{May} \bibinfo{year}{2024}), \bibinfo{pages}{926–938}.
\newblock
\showISSN{2150-8097}
\href{https://doi.org/10.14778/3641204.3641205}{doi:\nolinkurl{10.14778/3641204.3641205}}


\bibitem[Woltmann et~al\mbox{.}(2023)]%
        {woltmann2023fastgres}
\bibfield{author}{\bibinfo{person}{Lucas Woltmann}, \bibinfo{person}{Jerome Thiessat}, \bibinfo{person}{Claudio Hartmann}, \bibinfo{person}{Dirk Habich}, {and} \bibinfo{person}{Wolfgang Lehner}.} \bibinfo{year}{2023}\natexlab{}.
\newblock \showarticletitle{FASTgres: Making Learned Query Optimizer Hinting Effective}.
\newblock \bibinfo{journal}{\emph{Proc. VLDB Endow.}} \bibinfo{volume}{16}, \bibinfo{number}{11} (\bibinfo{date}{July} \bibinfo{year}{2023}), \bibinfo{pages}{3310–3322}.
\newblock
\showISSN{2150-8097}
\href{https://doi.org/10.14778/3611479.3611528}{doi:\nolinkurl{10.14778/3611479.3611528}}


\bibitem[Wu et~al\mbox{.}(2023)]%
        {wu2023factorjoin}
\bibfield{author}{\bibinfo{person}{Ziniu Wu}, \bibinfo{person}{Parimarjan Negi}, \bibinfo{person}{Mohammad Alizadeh}, \bibinfo{person}{Tim Kraska}, {and} \bibinfo{person}{Samuel Madden}.} \bibinfo{year}{2023}\natexlab{}.
\newblock \showarticletitle{FactorJoin: a new cardinality estimation framework for join queries}.
\newblock \bibinfo{journal}{\emph{Proceedings of the ACM on Management of Data}} \bibinfo{volume}{1}, \bibinfo{number}{1} (\bibinfo{year}{2023}), \bibinfo{pages}{1--27}.
\newblock


\bibitem[Wu et~al\mbox{.}(2020)]%
        {wu2020bayescard}
\bibfield{author}{\bibinfo{person}{Ziniu Wu}, \bibinfo{person}{Amir Shaikhha}, \bibinfo{person}{Rong Zhu}, \bibinfo{person}{Kai Zeng}, \bibinfo{person}{Yuxing Han}, {and} \bibinfo{person}{Jingren Zhou}.} \bibinfo{year}{2020}\natexlab{}.
\newblock \showarticletitle{Bayescard: Revitilizing bayesian frameworks for cardinality estimation}.
\newblock \bibinfo{journal}{\emph{arXiv preprint arXiv:2012.14743}} (\bibinfo{year}{2020}).
\newblock


\bibitem[Xu et~al\mbox{.}(2023)]%
        {xu2023coool}
\bibfield{author}{\bibinfo{person}{Xianghong Xu}, \bibinfo{person}{Zhibing Zhao}, \bibinfo{person}{Tieying Zhang}, \bibinfo{person}{Rong Kang}, \bibinfo{person}{Luming Sun}, {and} \bibinfo{person}{Jianjun Chen}.} \bibinfo{year}{2023}\natexlab{}.
\newblock \bibinfo{title}{COOOL: A Learning-To-Rank Approach for SQL Hint Recommendations}.
\newblock
\showeprint[arxiv]{2304.04407}~[cs.DB]
\urldef\tempurl%
\url{https://arxiv.org/abs/2304.04407}
\showURL{%
\tempurl}


\bibitem[Yang et~al\mbox{.}(2022)]%
        {Balsa_Yang2022}
\bibfield{author}{\bibinfo{person}{Zongheng Yang}, \bibinfo{person}{Wei~Lin Chiang}, \bibinfo{person}{Sifei Luan}, \bibinfo{person}{Gautam Mittal}, \bibinfo{person}{Michael Luo}, {and} \bibinfo{person}{Ion Stoica}.} \bibinfo{year}{2022}\natexlab{}.
\newblock \showarticletitle{Balsa: Learning a Query Optimizer Without Expert Demonstrations}.
\newblock \bibinfo{journal}{\emph{Proceedings of the ACM SIGMOD International Conference on Management of Data}} (\bibinfo{date}{6} \bibinfo{year}{2022}), \bibinfo{pages}{931--944}.
\newblock
\showISBNx{9781450392495}
\showISSN{07308078}
\href{https://doi.org/10.1145/3514221.3517885}{doi:\nolinkurl{10.1145/3514221.3517885}}


\bibitem[Yang et~al\mbox{.}(2020)]%
        {NeuroCard_yang2020neurocard}
\bibfield{author}{\bibinfo{person}{Zongheng Yang}, \bibinfo{person}{Amog Kamsetty}, \bibinfo{person}{Sifei Luan}, \bibinfo{person}{Eric Liang}, \bibinfo{person}{Yan Duan}, \bibinfo{person}{Xi Chen}, {and} \bibinfo{person}{Ion Stoica}.} \bibinfo{year}{2020}\natexlab{}.
\newblock \showarticletitle{Neurocard: One cardinality estimator for all tables}.
\newblock \bibinfo{journal}{\emph{arXiv preprint arXiv:2006.08109}} (\bibinfo{year}{2020}).
\newblock


\bibitem[Yu et~al\mbox{.}(2022)]%
        {yu2022hybridqo}
\bibfield{author}{\bibinfo{person}{Xiang Yu}, \bibinfo{person}{Chengliang Chai}, \bibinfo{person}{Guoliang Li}, {and} \bibinfo{person}{Jiabin Liu}.} \bibinfo{year}{2022}\natexlab{}.
\newblock \showarticletitle{Cost-based or learning-based? A hybrid query optimizer for query plan selection}.
\newblock \bibinfo{journal}{\emph{Proceedings of the VLDB Endowment}} \bibinfo{volume}{15}, \bibinfo{number}{13} (\bibinfo{year}{2022}), \bibinfo{pages}{3924--3936}.
\newblock


\bibitem[Yu et~al\mbox{.}(2020)]%
        {RTOS_yu2020}
\bibfield{author}{\bibinfo{person}{Xiang Yu}, \bibinfo{person}{Guoliang Li}, \bibinfo{person}{Chengliang Chai}, {and} \bibinfo{person}{Nan Tang}.} \bibinfo{year}{2020}\natexlab{}.
\newblock \showarticletitle{Reinforcement learning with tree-lstm for join order selection}. In \bibinfo{booktitle}{\emph{2020 IEEE 36th International Conference on Data Engineering (ICDE)}}. IEEE, \bibinfo{pages}{1297--1308}.
\newblock


\bibitem[Zhao et~al\mbox{.}(2022)]%
        {zhao2022queryformer}
\bibfield{author}{\bibinfo{person}{Yue Zhao}, \bibinfo{person}{Gao Cong}, \bibinfo{person}{Jiachen Shi}, {and} \bibinfo{person}{Chunyan Miao}.} \bibinfo{year}{2022}\natexlab{}.
\newblock \showarticletitle{Queryformer: A tree transformer model for query plan representation}.
\newblock \bibinfo{journal}{\emph{Proceedings of the VLDB Endowment}} \bibinfo{volume}{15}, \bibinfo{number}{8} (\bibinfo{year}{2022}), \bibinfo{pages}{1658--1670}.
\newblock


\bibitem[Zhu et~al\mbox{.}(2023)]%
        {zhu2023lero}
\bibfield{author}{\bibinfo{person}{Rong Zhu}, \bibinfo{person}{Wei Chen}, \bibinfo{person}{Bolin Ding}, \bibinfo{person}{Xingguang Chen}, \bibinfo{person}{Andreas Pfadler}, \bibinfo{person}{Ziniu Wu}, {and} \bibinfo{person}{Jingren Zhou}.} \bibinfo{year}{2023}\natexlab{}.
\newblock \showarticletitle{Lero: A learning-to-rank query optimizer}.
\newblock \bibinfo{journal}{\emph{Proceedings of the VLDB Endowment}} \bibinfo{volume}{16}, \bibinfo{number}{6} (\bibinfo{year}{2023}), \bibinfo{pages}{1466--1479}.
\newblock


\bibitem[Zhu et~al\mbox{.}(2024)]%
        {zhu2024pilotscope}
\bibfield{author}{\bibinfo{person}{Rong Zhu}, \bibinfo{person}{Lianggui Weng}, \bibinfo{person}{Wenqing Wei}, \bibinfo{person}{Di Wu}, \bibinfo{person}{Jiazhen Peng}, \bibinfo{person}{Yifan Wang}, \bibinfo{person}{Bolin Ding}, \bibinfo{person}{Defu Lian}, \bibinfo{person}{Bolong Zheng}, {and} \bibinfo{person}{Jingren Zhou}.} \bibinfo{year}{2024}\natexlab{}.
\newblock \showarticletitle{PilotScope: Steering Databases with Machine Learning Drivers}.
\newblock \bibinfo{journal}{\emph{Proc. VLDB Endow.}} \bibinfo{volume}{17}, \bibinfo{number}{5} (\bibinfo{date}{May} \bibinfo{year}{2024}), \bibinfo{pages}{980–993}.
\newblock
\showISSN{2150-8097}
\href{https://doi.org/10.14778/3641204.3641209}{doi:\nolinkurl{10.14778/3641204.3641209}}


\bibitem[Zhu et~al\mbox{.}(2020)]%
        {zhu2020flat}
\bibfield{author}{\bibinfo{person}{Rong Zhu}, \bibinfo{person}{Ziniu Wu}, \bibinfo{person}{Yuxing Han}, \bibinfo{person}{Kai Zeng}, \bibinfo{person}{Andreas Pfadler}, \bibinfo{person}{Zhengping Qian}, \bibinfo{person}{Jingren Zhou}, {and} \bibinfo{person}{Bin Cui}.} \bibinfo{year}{2020}\natexlab{}.
\newblock \showarticletitle{FLAT: fast, lightweight and accurate method for cardinality estimation}.
\newblock \bibinfo{journal}{\emph{arXiv preprint arXiv:2011.09022}} (\bibinfo{year}{2020}).
\newblock


\end{thebibliography}

\end{document}